%% file: elastic_1km_epjc.tex
\documentclass[pdftex,twocolumn,epjc3]{svjour3}

\RequirePackage[T1]{fontenc}

\RequirePackage{graphicx}
\RequirePackage{mathptmx}      
\RequirePackage{flushend}
\RequirePackage[numbers,sort&compress]{natbib}
\RequirePackage[colorlinks,citecolor=blue,urlcolor=blue,linkcolor=blue]{hyperref}
\RequirePackage{amsmath}
\RequirePackage{amssymb}
\RequirePackage{enumitem}

\def\d{{\rm d}}
\def\un#1{\,{\rm #1}}
\def\ung#1{\quad[{\rm #1}]}
\def\unt#1{[{\rm #1}]}
\def\e{{\rm e}}
\def\I{{\rm i}}
\def\T{{\rm T}}
\def\mat#1{\tens{#1}}
\def\etal{et al.}

\setbox123\hbox{$0$}
\setbox124\hbox{$.$}
\def\S{\hbox to\wd123{\hss}}
\def\.{\hbox to\wd124{\hss}}

\def\Name#1{#1, }
\def\Review#1#2#3#4{{\it #1} {\bf #2} (#3) #4}

\def\Break{\break}

\journalname{Eur. Phys. J. C}

\begin{document}

\title{Measurement of Elastic pp Scattering at $\sqrt{\hbox{s}} = \hbox{8}$\,TeV in the Coulomb-Nuclear Interference Region -- Determination of the $\mathbf{\rho}$-Parameter and the Total Cross-Section}

\titlerunning{Meas.~of Elastic pp Scatt.~at $\sqrt{\hbox{s}} = \hbox{8}$\,TeV in the CNI Region -- Determination of $\rho$ and $\sigma_{\rm tot}$}

\input authorlist_epjc

\authorrunning{The TOTEM Collaboration}

\date{Manuscript date: \today}

\maketitle

\begin{abstract}
\input abstract
\end{abstract}


\input introduction.tex

\input experimental_apparatus.tex

\input beam_optics.tex

\input data_taking.tex

\input differential_cross_section.tex

\input coulomb.tex

\input conclusions.tex

\begin{acknowledgements}
\input acknowledgements.tex
\end{acknowledgements}


\end{document}

%% file: authorlist_epjc.tex

\newif\ifFirstAuthor
\FirstAuthortrue

\def\AddAuthor#1#2#3#4{%
	\def\PriAf{#2}%
	\def\SecAf{#3}%
	\def\ExtAf{#4}%
	\def\empty{}%
	\ifFirstAuthor
		\FirstAuthorfalse
	\else
		\and
	\fi
	\ifx\PriAf\empty
		#1\thanksref{#4}%
	\else
		\ifx\SecAf\empty
			\ifx\ExtAf\empty
				#1\thanksref{#2}%
			\else
				#1\thanksref{#2,#4}%
			\fi
		\else
			\ifx\ExtAf\empty
				#1\thanksref{#2,#3}%
			\else
				#1\thanksref{#2,#3,#4}%
			\fi
		\fi
	\fi
}



\newif\ifFirstInstitute
\FirstInstitutetrue

\def\AddInstitute#1#2{%
	\ifFirstInstitute
		\FirstInstitutefalse
	\else
		\and
	\fi
	#2\label{#1}%
}


\def\AddExternalInstitute#1#2{%
	\thankstext{#1}{#2}
}


\input authorlist_data

\author{%
	The TOTEM Collaboration\\
	\DeclareAuthors
}

\DeclareExternalInstitutes

\institute{%
	\DeclareInstitutes
}

%% file: authorlist_data.tex
\def\DeclareAuthors{%
	\AddAuthor{G.~Antchev}{}{}{a}%
	\AddAuthor{P.~Aspell}{9}{}{}%
	\AddAuthor{I.~Atanassov}{}{}{a}%
	\AddAuthor{V.~Avati}{8}{9}{}%
	\AddAuthor{J.~Baechler}{9}{}{}%
	\AddAuthor{V.~Berardi}{5b}{5a}{}%
	\AddAuthor{M.~Berretti}{9}{7b}{}%
	\AddAuthor{E.~Bossini}{7b}{}{}%
	\AddAuthor{U.~Bottigli}{7b}{}{}%
	\AddAuthor{M.~Bozzo}{6a}{6b}{}%
	\AddAuthor{P.~Broul\'{i}m}{1a}{}{}%
	\AddAuthor{H.~Burkhardt}{9}{}{}%
	\AddAuthor{A.~Buzzo}{6a}{}{}%
	\AddAuthor{F.~S.~Cafagna}{5a}{}{}%
	\AddAuthor{C.~E.~Campanella}{5c}{5a}{}%
	\AddAuthor{M.~G.~Catanesi}{5a}{}{}%
	\AddAuthor{M.~Csan\'{a}d}{4a}{}{b}%
	\AddAuthor{T.~Cs\"{o}rg\H{o}}{4a}{4b}{}%
	\AddAuthor{M.~Deile}{9}{}{}%
	\AddAuthor{F.~De~Leonardis}{5c}{5a}{}%
	\AddAuthor{A.~D'Orazio}{5c}{5a}{}%
	\AddAuthor{M.~Doubek}{1c}{}{}%
	\AddAuthor{K.~Eggert}{10}{}{}%
	\AddAuthor{V.~Eremin}{}{}{e}%
	\AddAuthor{F.~Ferro}{6a}{}{}%
	\AddAuthor{A. Fiergolski}{5a}{}{d}%
	\AddAuthor{F.~Garcia}{3a}{}{}%
	\AddAuthor{V.~Georgiev}{1a}{}{}%
	\AddAuthor{S.~Giani}{9}{}{}%
	\AddAuthor{L.~Grzanka}{8}{}{c}%
	\AddAuthor{C.~Guaragnella}{5c}{5a}{}%
	\AddAuthor{J.~Hammerbauer}{1a}{}{}%
	\AddAuthor{J.~Heino}{3a}{}{}%
	\AddAuthor{A.~Karev}{9}{}{}%
	\AddAuthor{J.~Ka\v{s}par}{1b}{9}{}%
	\AddAuthor{J.~Kopal}{1b}{}{}%
	\AddAuthor{V.~Kundr\'{a}t}{1b}{}{}%
	\AddAuthor{S.~Lami}{7a}{}{}%
	\AddAuthor{G.~Latino}{7b}{}{}%
	\AddAuthor{R.~Lauhakangas}{3a}{}{}%
	\AddAuthor{R.~Linhart}{1a}{}{}%
	\AddAuthor{E.~Lippmaa}{2}{}{+}%
	\AddAuthor{J.~Lippmaa}{2}{}{}%
	\AddAuthor{M.~V.~Lokaj\'{\i}\v{c}ek}{1b}{}{}%
	\AddAuthor{L.~Losurdo}{7b}{}{}%
	\AddAuthor{M.~Lo~Vetere}{6b}{6a}{+}%
	\AddAuthor{F.~Lucas~Rodr\'{i}guez}{9}{}{}%
	\AddAuthor{M.~Macr\'{\i}}{6a}{}{}%
	\AddAuthor{A.~Mercadante}{5a}{}{}%
	\AddAuthor{N.~Minafra}{9}{5b}{}%
	\AddAuthor{S.~Minutoli}{6a}{}{}%
	\AddAuthor{T.~Naaranoja}{3a}{3b}{}%
	\AddAuthor{F.~Nemes}{4a}{}{b}%
	\AddAuthor{H.~Niewiadomski}{10}{}{}%
	\AddAuthor{E.~Oliveri}{9}{}{}%
	\AddAuthor{F.~Oljemark}{3a}{3b}{}%
	\AddAuthor{R.~Orava}{3a}{3b}{}%
	\AddAuthor{M.~Oriunno}{}{}{f}%
	\AddAuthor{K.~\"{O}sterberg}{3a}{3b}{}%
	\AddAuthor{P.~Palazzi}{9}{}{}%
	\AddAuthor{L.~Palo\v{c}ko}{1a}{}{}%
	\AddAuthor{V.~Passaro}{5c}{5a}{}%
	\AddAuthor{Z.~Peroutka}{1a}{}{}%
	\AddAuthor{V.~Petruzzelli}{5c}{5a}{}%
	\AddAuthor{T.~Politi}{5c}{5a}{}%
	\AddAuthor{J.~Proch\'{a}zka}{1b}{}{}%
	\AddAuthor{F.~Prudenzano}{5c}{5a}{}%
	\AddAuthor{M.~Quinto}{5a}{5b}{}%
	\AddAuthor{E.~Radermacher}{9}{}{}%
	\AddAuthor{E.~Radicioni}{5a}{}{}%
	\AddAuthor{F.~Ravotti}{9}{}{}%
	\AddAuthor{S.~Redaelli}{9}{}{}%
	\AddAuthor{E.~Robutti}{6a}{}{}%
	\AddAuthor{L.~Ropelewski}{9}{}{}%
	\AddAuthor{G.~Ruggiero}{9}{}{}%
	\AddAuthor{H.~Saarikko}{3a}{3b}{}%
	\AddAuthor{B.~Salvachua}{9}{}{}%
	\AddAuthor{A.~Scribano}{7a}{}{}%
	\AddAuthor{J.~Smajek}{9}{}{}%
	\AddAuthor{W.~Snoeys}{9}{}{}%
	\AddAuthor{J.~Sziklai}{4a}{}{}%
	\AddAuthor{C.~Taylor}{10}{}{}%
	\AddAuthor{N.~Turini}{7b}{}{}%
	\AddAuthor{V.~Vacek}{1c}{}{}%
	\AddAuthor{G.~Valentino}{9}{}{}%
	\AddAuthor{J.~Welti}{3a}{3b}{}%
	\AddAuthor{J.~Wenninger}{9}{}{}%
	\AddAuthor{P.~Wyszkowski}{8}{}{}%
	\AddAuthor{K.~Zielinski}{8}{}{}%
}


\def\DeclareInstitutes{%
	\AddInstitute{1a}{University of West Bohemia, Pilsen, Czech Republic.}
	\AddInstitute{1b}{Institute of Physics of the Academy of Sciences of the Czech Republic, Prague, Czech Republic.}
	\AddInstitute{1c}{Czech Technical University, Prague, Czech Republic.}
	\AddInstitute{2}{National Institute of Chemical Physics and Biophysics NICPB, Tallinn, Estonia.}
	\AddInstitute{3a}{Helsinki Institute of Physics, Helsinki, Finland.}
	\AddInstitute{3b}{Department of Physics, University of Helsinki, Helsinki, Finland.}
	\AddInstitute{4a}{Wigner Research Centre for Physics, RMKI, Budapest, Hungary.}
	\AddInstitute{4b}{EKU KRC, Gy\"ongy\"os, Hungary.}
	\AddInstitute{5a}{INFN Sezione di Bari, Bari, Italy.}
	\AddInstitute{5b}{Dipartimento Interateneo di Fisica di Bari, Bari, Italy.}
	\AddInstitute{5c}{Dipartimento di Ingegneria Elettrica e dell'Informazione - Politecnico di Bari, Bari, Italy.}
	\AddInstitute{6a}{INFN Sezione di Genova, Genova, Italy.}
	\AddInstitute{6b}{Universit\`{a} degli Studi di Genova, Italy.}
	\AddInstitute{7a}{INFN Sezione di Pisa, Pisa, Italy.}
	\AddInstitute{7b}{Universit\`{a} degli Studi di Siena and Gruppo Collegato INFN di Siena, Siena, Italy.}
	\AddInstitute{8}{AGH University of Science and Technology, Krakow, Poland.}
	\AddInstitute{9}{CERN, Geneva, Switzerland.}
	\AddInstitute{10}{Case Western Reserve University, Dept.~of Physics, Cleveland, OH, USA.}
}

	
\def\DeclareExternalInstitutes{%
	\AddExternalInstitute{a}{INRNE-BAS, Institute for Nuclear Research and Nuclear Energy, Bulgarian Academy of Sciences, Sofia, Bulgaria.}
	\AddExternalInstitute{b}{Department of Atomic Physics, ELTE University, Budapest, Hungary.}
	\AddExternalInstitute{c}{Institute of Nuclear Physics, Polish Academy of Science, Krakow, Poland.}
	\AddExternalInstitute{d}{Warsaw University of Technology, Warsaw, Poland.}
	\AddExternalInstitute{e}{Ioffe Physical - Technical Institute of Russian Academy of Sciences, St.~Petersburg, Russian Federation.}
	\AddExternalInstitute{f}{SLAC National Accelerator Laboratory, Stanford CA, USA.}
	\AddExternalInstitute{+}{Deceased.}
}

%% file: abstract.tex
The TOTEM experiment at the CERN LHC has measured elastic proton-proton 
scattering at the centre-of-mass energy 
$\sqrt{s}=8\,$TeV and four-momentum transfers\Break
squared, $|t|$, from $6\times10^{-4}\,\rm GeV^{2}$ to 0.2\,GeV$^{2}$.
Near the\Break lower end of the $t$-interval the differential cross-section is 
sensitive to the 
interference between the hadronic and the electromagnetic scattering amplitudes.
This article presents the elastic cross-section measurement and the constraints it 
imposes on the functional forms of the modulus and
phase of the hadronic elastic amplitude. The data exclude the traditional 
Simplified West and Yennie interference formula that requires a constant 
phase and a purely exponential modulus of the hadronic amplitude. 
For parametrisations of the hadronic modulus with second- or third-order 
polynomials in the exponent, the data are compatible with hadronic phase 
functions giving either central or peripheral behaviour in the impact 
parameter picture of elastic scattering. In both cases, 
the $\rho$-parameter is found to be $0.12 \pm 0.03$. 
The results for the total hadronic 
cross-section are $\sigma_{\rm tot} = (102.9 \pm 2.3)$\,mb and 
$(103.0 \pm 2.3)$\,mb for central and peripheral phase formulations, 
respectively. Both are consistent with previous TOTEM measurements.

%% file: introduction.tex
\section{Introduction}
\label{sec:introduction}

Elastic scattering of protons is a process mediated by the strong and the electromagnetic interactions -- the weak interaction is commonly neglected since its carriers are heavy compared to the small momentum transfers, $|t|$, typical of elastic scattering. In this context, the strong interaction is traditionally called `nuclear' or `hadronic' and the electromagnetic one `Coulomb'. In quantum-theory description,\Break each of these interactions is described by a scattering amplitude, nuclear ${\cal A}^{\rm N}(t)$ and Coulomb ${\cal A}^{\rm C}(t)$. Moreover, the combined scattering amplitude receives a third contribution reflecting Feynman diagrams with both strong and electromagnetic exchanges. This term, together with the complex character of the scattering amplitudes, describes the effects of Coulomb-nuclear interference (CNI) in the differential cross-section. Since the Coulomb amplitude is known, measuring the CNI gives access to the phase of the nuclear amplitude, which is necessary for a complete understanding of the interaction but not directly observable in the pure hadronic differential cross-section. The CNI effect is most pronounced in the $t$-region where the two amplitudes have similar magnitudes, i.e.~-- for typical LHC centre-of-mass energies of a few TeV -- near $|t| \sim 5 \times 10^{-4}\,\rm GeV^{2}$. Thus the experimental sensitivity to the nuclear phase, $\arg {\cal A}^{\rm N}(t)$, is limited to a region at very small $|t|$, making difficult any conclusions on the 
functional form of the phase.

In the analyses of past experiments -- see e.g.~\cite{plb43,plb66,npb141,prl47,plb115,plb120,plb128,npb262}\Break (ISR), \cite{plb198,plb316} ($\rm S\bar{p}pS$), \cite{prl68} (Tevatron) --
a simplified interference formula was used. This so-called
Simplified West-Yennie (SWY) formula~\cite{wy68} is based on restrictive assumptions on the 
hadronic amplitude, implying in particular a purely exponential modulus and a 
constant phase for all $t$ (see the discussion in 
Section~\ref{sec:cni interference}).
As a representative quantity, the phase value at $t=0$, or equivalently
\begin{equation}
\label{eq:rho def}
\rho \equiv \cot \, \arg {\cal A}^{\rm N} (0) = \frac{\Re {\cal A}^{\rm N} (0)}{\Im {\cal A}^{\rm N} (0)}
\end{equation}
was traditionally quoted. 
An interesting aspect of $\rho$ is its predictive power in extrapolating the total cross-section to higher centre-of-mass energies via dispersion 
relations~\cite{dremin-dispersion}. 

The present article discusses the first measurement of elastic scattering in the CNI region at the CERN LHC by the TOTEM experiment. The data have been collected at $\sqrt{s} = 8\,$TeV with a special beam optics ($\beta^{*}=1000\,$m) and cover a $|t|$-interval from $6\times10^{-4}\,\rm GeV^{2}$ to 0.2\,GeV$^{2}$, extending well into the interference region. In order to strengthen the statistical power and thus enable a cleaner identification of the interference effects, the analysis also exploits another, complementary data set with higher statistics~\cite{8tev-90m}, taken at the same energy, but with different beam optics ($\beta^{*}=90\,$m), and thus covering a different $|t|$-range: $0.027 < |t| < 0.2\un{GeV^2}$. The isolated analysis of the latter data set has excluded a purely exponential behaviour of the observed elastic cross-section with more than $7\un{\sigma}$ confidence. The new data in the CNI region allow to study the source of the non-exponentiality: nuclear component, CNI effects or both. In order to explore the full spectrum of possibilities, an interference formula without the limitations of SWY is needed. In the present study the more general and complex interference formulae of Cahn \cite{cahn82} and Kundr\'{a}t-Lokaj\'{\i}\v{c}ek (KL) \cite{kl94} are used, offering much more freedom for the choice of the theoretically unknown functional forms of the hadronic modulus and phase. Since the data cannot unambiguously determine all functional forms and their parameters, the results of this study, still representatively expressed in terms of $\rho$, become conditional to the choice of the model describing the hadronic amplitude. This choice has implications on the behaviour of the interaction in impact parameter space. In particular, the functional form of the hadronic phase at small $|t|$ determines whether elastic collisions occur predominantly at small or large impact parameters (centrality vs.~peripherality). It will be shown that both options are compatible with the data, thus the central picture still prevalent in theoretical models is not a necessity.

Section~\ref{sec:exp apparatus} of this article outlines the experimental setup used for the measurement. The properties of the special\Break beam optics are described in Section~\ref{sec:beam optics}. Section~\ref{sec:data taking} gives details of the data-taking conditions. The data analysis and reconstruction of the differential cross-section are described in Section~\ref{sec:differential cross-section}. Section~\ref{sec:coulomb} presents the study of the Coulomb-nuclear interference together with the functional form of the hadronic amplitude. The values of $\rho$ and $\sigma_{\rm tot}$ are determined.

%% file: experimental_apparatus.tex
\section{Experimental Apparatus}
\label{sec:exp apparatus}

\begin{figure*}
\begin{center}
\includegraphics{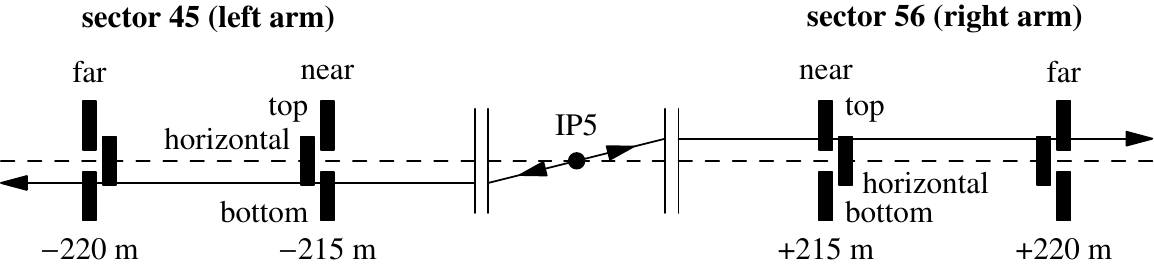}
\hfil
\includegraphics{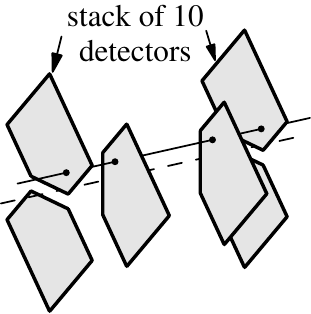}
\caption{%
Left: schematic view of the RP stations on both sides of IP5 with two proton tracks from an elastic event. Right: Schematic view of the silicon detector positions in an RP station with a track traversing the overlap zone between top and horizontal detectors, providing detector alignment information. 
}
\label{fig:rpsketch}
\end{center}
\end{figure*}

The TOTEM experiment, located at the LHC Interaction Point (IP) 5 together with
the CMS experiment, is dedicated to the measurement of the total 
cross-section, elastic scattering
and diffractive processes. The experimental
apparatus, symmetric with respect to the IP, is 
composed of a forward proton spectrometer (Roman Pots, RPs) and the 
forward tracking telescopes T1 and T2. 
A complete description of the TOTEM detector instrumentation 
and its performance is given in~\cite{totem-jinst} and~\cite{totem-ijmp}. 
The data analysed here come from the RPs only. An RP is a movable beam-pipe
insertion capable of approaching the LHC beam to a distance of less than a millimetre, in 
order to detect protons with scattering angles of only a few microradians. 
The proton spectrometer is organised in two RP stations: one on the left side of the IP 
(LHC sector 45) and one on the right (LHC sector 56), see Figure~\ref{fig:rpsketch} (left).
Each RP station, located between 215 and 220\,m from the IP, is composed of two 
units: ``near'' (215\,m from the IP) and ``far'' (220\,m). 
A unit consists of 3 RPs, one
approaching the outgoing beam from the top, one from the bottom, and one 
horizontally.
Each RP houses a stack of 5 ``U'' and 5 ``V'' silicon strip detectors, where ``U'' and ``V''
refer to two mutually perpendicular strip orientations. The sensors were designed with the
specific objective of reducing the insensitive area at the edge facing the beam
to only a few tens of micrometers. Due to the 5\,m long lever arm 
between the near and the far RP units 
the local track angles can be reconstructed
with a precision of about $10\,\mu$rad. A high trigger efficiency
($> 99$\%) is achieved by using all RPs independently. 
Since elastic scattering events consist of two collinear protons emitted in 
opposite directions, the detected events can have two topologies, called 
diagonals: 45 bottom -- 56 top and 45 top -- 56 bottom.

This article uses a reference frame where $x$ denotes the horizontal axis (pointing out of the LHC ring), $y$ the vertical axis (pointing against gravity) and $z$ the beam axis (in the clockwise direction).

%% file: beam_optics.tex
\section{Beam Optics}
\label{sec:beam optics}

The beam optics relates the proton kinematical states at the IP and at the RP location. A proton emerging from the interaction vertex $(x^*$, $y^*)$ at the angle $(\theta_x^*,\theta_y^*)$ (relative to the $z$ axis) and with momentum $p\,(1+\xi)$, where $p$ is the nominal initial-state proton momentum, is transported along the outgoing beam through the LHC magnets. It arrives at the RPs in the transverse position
\begin{equation}
\label{eq:prot trans}
	\begin{aligned}
		x(z_{\rm RP}) =& L_x(z_{\rm RP})\, \theta_x^*\ +\ v_x(z_{\rm RP})\, x^*\ +\ D_x(z_{\rm RP})\, \xi\ ,\cr
		y(z_{\rm RP}) =& L_y(z_{\rm RP})\, \theta_y^*\ +\ v_y(z_{\rm RP})\, y^*\ +\ D_y(z_{\rm RP})\, \xi \quad
	\end{aligned}
\end{equation}
relative to the beam centre. This position is determined by the optical functions, characterising the transport of protons in the beam line and controlled via 
the LHC magnet currents.
The effective length $L_{x,y}(z)$, magnification $v_{x,y}(z)$ and dispersion $D_{x,y}(z)$ quantify the sensitivity of the measured proton position to the 
scattering angle, vertex position and momentum loss, respectively.
Note that for elastic collisions the dispersion terms $D\,\xi$ can be ignored because the protons do not lose any momentum. The values of $\xi$ only account for the initial state momentum offset and variations, see Section 4 in~\cite{8tev-90m}. Due to the collinearity of the two elastically scattered protons and the symmetry of the optics, the impact of $D\,\xi$ on the reconstructed scattering angles is negligible compared to other uncertainties.

The data for the analysis presented here have been taken with a new, special optics, conventionally labelled by the value of the $\beta$-function at the interaction point,\Break $\beta^{*} = 1000\,$m, and specifically developed for measuring low-$|t|$ elastic scattering. It maximises the vertical effective\Break length $L_{y}$ at the RP position $z = 220\un{m}$ and minimises the vertical magnification $|v_{y}|$ at $z = 220\,$m (Table~\ref{tab:optics}). This configuration is called
``parallel-to-point focussing'' because all protons with the same angle in the IP are focussed on one point in the RP at 220\,m. It optimises the sensitivity to the vertical projection of the scattering angle -- and hence to $|t|$ -- while minimising the influence of the vertex position. 
In the horizontal projection the parallel-to-point focussing condition is not fulfilled, but -- unlike in the $\beta^{*} = 90\,$m optics used for previous measurements~\cite{epl96,epl101-el,epl101-tot,prl111} -- the effective length $L_{x}$ at $z = 220\,$m is non-zero, which reduces the uncertainty in the horizontal component of the scattering angle.

\begin{table}
\caption{
Optical functions for elastic proton transport for the $\beta^{*} = 1000\,$m optics. The values refer to the right arm, for the left one they are very similar.
}
\label{tab:optics}
\begin{center}
\vskip-3mm
\begin{tabular}{ccccc}\hline
RP unit & $L_x$ & $v_x$ & $L_y$ & $v_y$ \cr\hline
near & $59.37\un{m}$  & $-0.867$ & $255.87\un{m}$ & $0.003$ \cr
far  & $45.89\un{m}$ & $-0.761$ & $284.62\un{m}$ & $-0.017$ \cr
\hline
\end{tabular}
\end{center}
\end{table}

%% file: data_taking.tex
\section{Data Taking}
\label{sec:data taking}

The results reported here are based on data taken in October 2012 
during a dedicated LHC proton fill (3216)
with the special beam properties described in the previous section.

\begin{figure*}
\begin{center}
\includegraphics{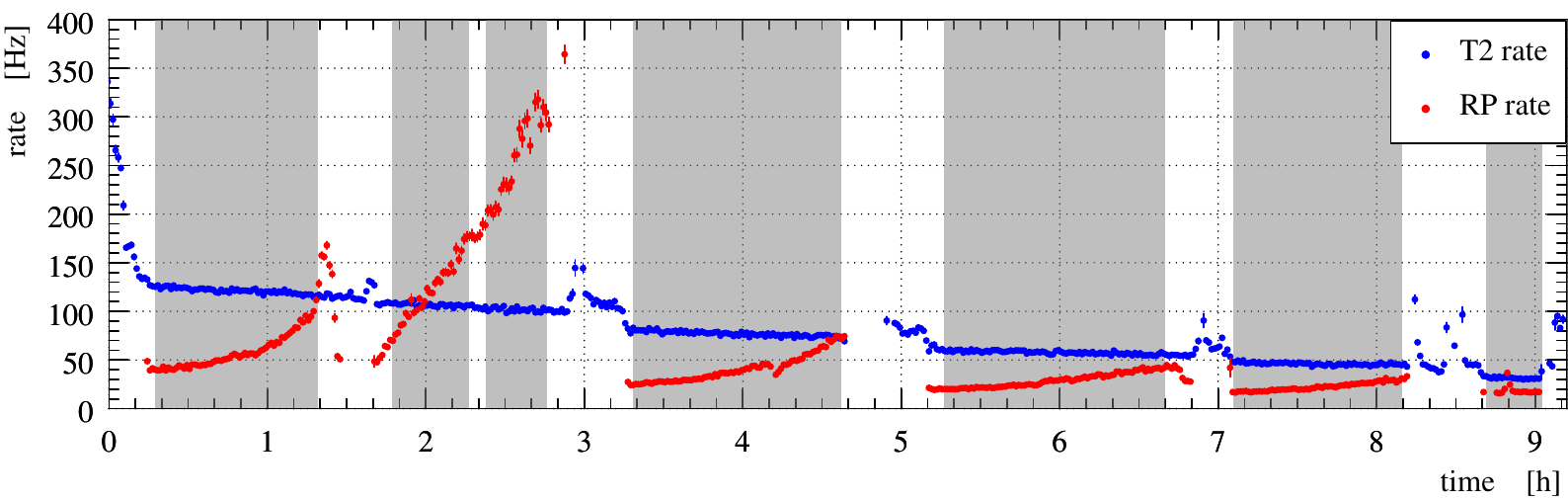}
\caption{%
Trigger rates as a function of time from the beginning of the run (24 October 2012, 23:00 h). The T2 rate (blue) is roughly proportional to luminosity, while the RP rate (red) is in addition sensitive to beam-halo level. The yellow bands represent periods of uninterrupted data taking.
}
\label{fig:overview}
\end{center}
\end{figure*}

The vertical RPs approached the beam centre to only 3 times the beam width, $\sigma_{y}$, resulting in an acceptance for $|t|$-values down to $6 \times 10^{-4}\,\rm GeV^{2}$. The exceptionally close distance was possible due to the low beam intensity in this special beam operation: each beam contained only two colliding bunches and one non-colliding bunch for background monitoring, each with $10^{11}$ protons. A novel collimation strategy was applied to keep the beam halo background under control. As a first step, the primary 
collimators (TCP) in the LHC betatron cleaning insertion (point 7) scraped the beam down to $2\,\sigma_{y}$; then the collimators were retracted to $2.5\,\sigma_{y}$, thus creating a $0.5\,\sigma_{y}$ gap between
the beam edge and the collimator jaws. With the halo strongly suppressed 
and no collimator producing showers by touching the beam, the RPs at 
$3\,\sigma_{y}$ were operated in a background-depleted environment for about one 
hour until the beam-to-collimator gap was refilled by diffusion, as 
diagnosed by the increasing RP trigger rate (Figure~\ref{fig:overview}). When the background conditions
had deteriorated to an unacceptable level, the beam cleaning procedure was repeated, again followed by a quiet data-taking period. The beam cleaning at $1.5\un{h}$ from the beginning of the run employed only vertical collimators and led to a quickly increasing background rate, see Figure~\ref{fig:overview}. Therefore, the following beam cleaning operations were also performed in the horizontal plane. Altogether there were 6 beam cleaning interventions until the luminosity had decreased from initially $1.8\times10^{27}\,\rm cm^{-2}s^{-1}$ to 
$0.4\times10^{27}\,\rm cm^{-2}s^{-1}$
at which point the data yield was considered as too low. 
During the 9 hour long fill, an integrated luminosity of $20\,\rm \mu b^{-1}$ 
was accumulated in 6 data sets corresponding to the calm periods 
between the cleaning operations. 

Due to an anti-collision protection system, the top and the bottom pots of a 
vertical RP unit could not approach each other close enough to be both at a 
distance of $3\,\sigma_{y} = 780\,\mu$m from the beam centre. Therefore a 
configuration with one RP diagonal (45 top -- 56 bottom) at $3\,\sigma_{y}$ (``close diagonal'') and the other (45 bottom -- 56 top) at 
$10\,\sigma_{y}$ (``distant diagonal'') was chosen. The distant diagonal provides a systematic comparison at larger $|t|$-values.
The horizontal RPs were only needed for the data-based alignment and therefore placed at a safe distance of $10\,\sigma_{x} \approx 7.5$\,mm, close enough to have an overlap with the vertical RPs (Figure~\ref{fig:rpsketch}, right).

The events collected were triggered by a logical \textit{OR} of: inelastic 
trigger (at least one charged particle in either arm of T2), double-arm proton trigger 
(coincidence of any RP left of IP5 and any RP right of IP5) and zero-bias trigger (random bunch crossings) for calibration purposes.

In the close and distant diagonals a total of 190k and 162k elastic event candidates have been tagged, respectively.

%% file: differential_cross_section.tex
\section{Differential Cross-Section}
\label{sec:differential cross-section}

The analysis method is very similar to the previously published one \cite{8tev-90m}. Section~\ref{sec:event analysis} covers all aspects related to the reconstruction of a single event. Section~\ref{sec:diff cs} describes the steps of transforming a raw $t$-distribution into the differential cross-section. The $t$-distributions for the two diagonals are analysed separately. After comparison (Section~\ref{sec:cross checks}) they are finally merged (Section~\ref{sec:final data merging}).

\subsection{Event Analysis}
\label{sec:event analysis}

The event kinematics are determined from the coordinates of track hits in the RPs after proper alignment (see Sec.~\ref{sec:alignment}) using the LHC optics (see Sec.~\ref{sec:optics}).


\subsubsection{Kinematics Reconstruction}
\label{sec:kinematics}

For each event candidate 
the scattering angles and vertex positions of both protons (one per arm) are first determined separately by inverting the proton transport equation~(\ref{eq:prot trans}), assuming $\xi = 0$:
\begin{equation}
\label{eq:kin 1a}
	\begin{aligned}
		\theta_x^{*\rm L,R} =& {v_x^{\rm N} x^{\rm F} - v_x^{\rm F} x^{\rm N}\over v_x^{\rm N} L_x^{\rm F} - v_x^{\rm F} L_x^{\rm N}}\ ,\quad
			\theta_y^{*\rm L,R} = {1\over 2} \left( {y^{\rm N}\over L_y^{\rm N}} + {y^{\rm F}\over L_y^{\rm F}} \right)\ ,\cr
		x^{*\rm L,R} =& {L_x^{\rm N} x^{\rm F} - L_x^{\rm F} x^{\rm N}\over L_x^{\rm N} v_x^{\rm F} - L_x^{\rm F} v_x^{\rm N}}\ ,\cr
	\end{aligned}
\end{equation}
where the N and F superscripts refer to the near and far units, L and R to the left and right arm, respectively. 
This one-arm reconstruction is used for tagging elastic events, where the left and right arm protons are compared.

Once a proton pair has been selected, all four RPs are used to reconstruct the kinematics of the event, optimising the angular resolution (see Section \ref{sec:resolution}):
\begin{equation}
\label{eq:kin 2a}
		\theta_x^* = {
				\sum {v_x^i}^2 \sum L_x^i x^i - \sum L_x^i v_x^i \sum v_x^i x^i
				\over
				\sum {v_x^i}^2 \sum {L_x^i}^2 - \sum L_x^i v_x^i \sum v_x^i L_x^i
			}\ ,\qquad
		\theta_y^* = {1\over 4} \sum {y^i\over L_y^i}\ ,
\end{equation}
where the sums run over the superscript $i$ representing the four RPs of a diagonal.

Eventually, the scattering angle, $\theta^*$, and the\Break four-momentum transfer squared, $t$, are calculated:

\begin{equation}
\label{eq:th t}
\theta^* = \sqrt{{\theta_x^*}^2 + {\theta_y^*}^2}\ ,\qquad t = - p^2 ({\theta_x^*}^2 + {\theta_y^*}^2)\ ,
\end{equation}
where $p$ denotes the beam momentum.


\subsubsection{Alignment}
\label{sec:alignment}

TOTEM's usual three-stage procedure~\cite{totem-ijmp} for correcting the detector positions and rotation angles  
has been applied: a beam-based alignment prior to the run followed by two offline methods. First, track-based alignment for relative positions among RPs, and second, alignment with elastic events for absolute position with respect to the beam -- repeated in 15 minutes time intervals to check for possible beam movements.

\begin{figure}
\begin{center}
\includegraphics{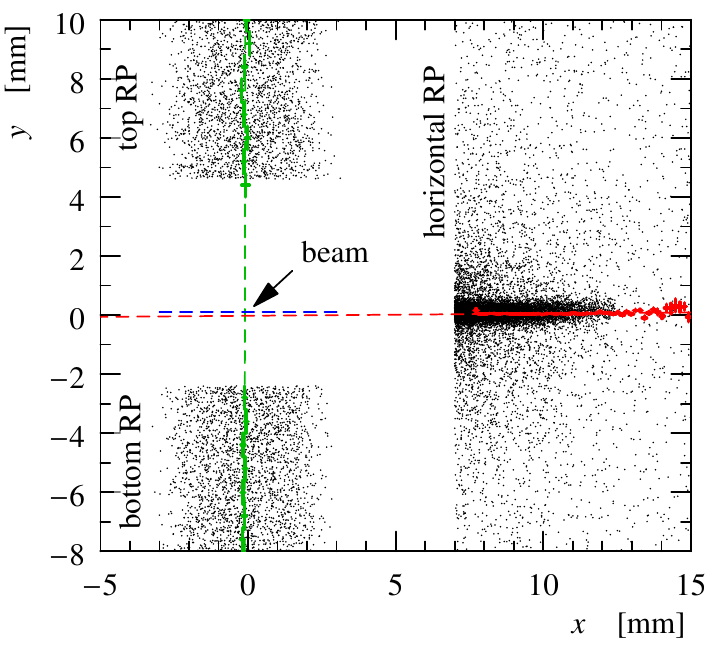}
\caption{%
Hit scatter plot in the right-far unit, corresponding to a period of 15 minutes. The black dots represent track hits in the vertical and horizontal RPs. Track hits close to the sensor edges are removed because of possible bias due to acceptance effects.
The green histogram shows the horizontal profile of hits in the vertical RPs, the dashed green line interpolates the profiles between the top and bottom RPs. Similarly, the red histogram gives the vertical profile of the hits in the horizontal RP and the dashed red line its extrapolation to the beam region. The blue dashed line indicates the vertical centre of symmetry of the hits in the vertical RPs (see \cite{totem-ijmp} for more details). The crossing of the dashed lines represents the position the beam centre (the two vertical-alignment results are averaged).
}
\label{fig:align meth}
\end{center}
\end{figure}

The offline procedure has been extended further to improve the vertical alignment. The new steps exploit the fact that elastic events with their two collinear protons relate the alignments in the left and right arm with an uncertainty of $20\un{\mu m}$. Furthermore, the horizontal RPs in the right arm recorded a hit distribution usable for vertical alignment in addition to the standard technique based on the vertical RPs, see Figure~\ref{fig:align meth}.

Exploiting all the methods, the alignment uncertainties have been estimated to $30\un{\mu m}$ (horizontal shift), $70\un{\mu m}$ (vertical shift) and $2\un{m rad}$ (rotation about the beam axis). Propagating them through Eq.~(\ref{eq:kin 2a}) to reconstructed scattering angles yields $0.28\un{\mu rad}$ ($0.19\un{\mu rad}$) for the horizontal (vertical) angle. RP rotations induce a bias in the reconstructed horizontal scattering angle:
\begin{equation}
\label{eq:alig rot bias}
	\theta_x^* \rightarrow \theta_x^* + c \theta_y^*\ ,
\end{equation}
where the proportionality constant $c$ has a mean of 0 and a standard deviation of $0.005$.


\subsubsection{Optics}
\label{sec:optics}
It is crucial to know with high precision the LHC beam optics between IP5 and the RPs, i.e. the behaviour of the spectrometer composed of the various magnetic elements.
The optics calibration has been applied as described in~\cite{totem-optics}. This method uses RP observables to determine fine corrections to the optical functions presented in Eq.~(\ref{eq:prot trans}).

The residual errors induce a bias in the reconstructed scattering angles:
\begin{equation}
\label{eq:opt bias}
	\theta_x^* \rightarrow (1 + d_x)\, \theta_x^*\ ,\qquad
	\theta_y^* \rightarrow (1 + d_y)\, \theta_y^*\ .
\end{equation}
For the two-arm reconstruction, Eq.~(\ref{eq:kin 2a}), the biases $d_x$ and $d_y$ have uncertainties of $0.34\un{\%}$ and $0.25\un{\%}$, respectively, and a correlation factor of $-0.89$. These estimates include the effects of magnet harmonics. To evaluate the impact on the $t$-distribution, it is convenient to decompose the correlated biases $d_x$ and $d_y$ into eigenvectors of the covariance matrix:
\begin{equation}
\label{eq:opt bias modes}
\begin{pmatrix} d_x\cr d_y \end{pmatrix} =
	\eta_1 \underbrace{\begin{pmatrix} +0.338\un{\%} \cr -0.234\un{\%} \end{pmatrix}}_{\rm mode\ 1}
	\ +\ \eta_2 \underbrace{\begin{pmatrix} -0.053\un{\%} \cr -0.076\un{\%} \end{pmatrix}}_{\rm mode\ 2}
\end{equation}
normalised such that the factors $\eta_{1,2}$ have unit variance.



\subsubsection{Resolution}
\label{sec:resolution}

\begin{figure}
\begin{center}
\includegraphics{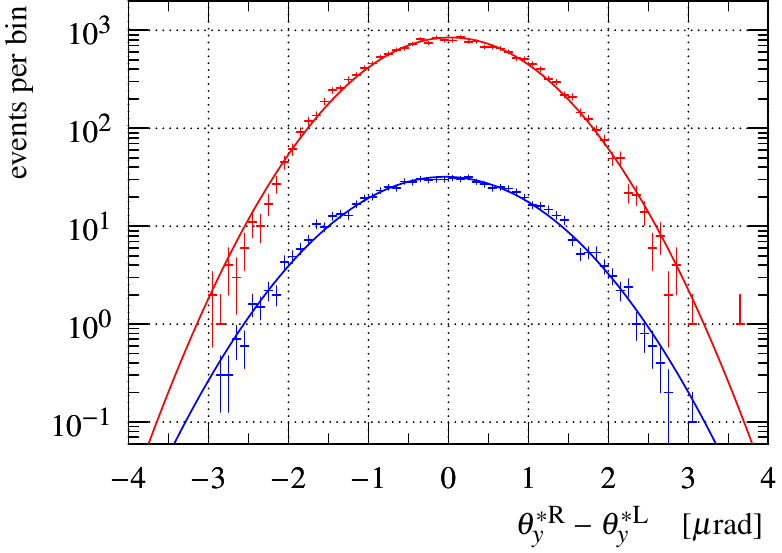}
\caption{%
Difference between vertical scattering angles reconstructed in the right and left arm, for the diagonal 45 top - 56 bottom. Red: data from run start ($0$ to $1\un{h}$ from the beginning of the run). Blue: data from run end ($7$ to $8\un{h}$), scaled by $0.1$. The solid lines represent Gaussian fits.
}
\label{fig:beam div vert}
\end{center}
\end{figure}

\begin{figure}
\begin{center}
\includegraphics{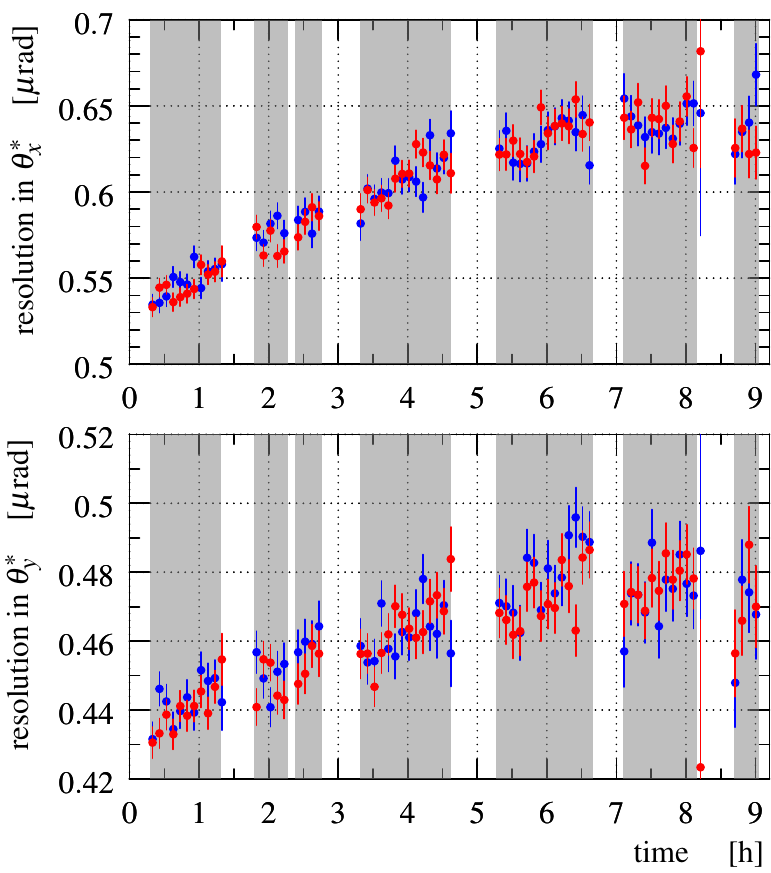}
\caption{%
Angular resolution for the two-arm reconstruction, Eq.~(\ref{eq:kin 2a}), as a function of time (from the beginning of the run). The blue (red) dots correspond to the diagonal 45 bottom -- 56 top (45 top -- 56 bottom). The grey bands indicate regions of uninterrupted data-taking, whereas in the remaining periods the beam cleaning procedure described in Section~\ref{sec:data taking} was performed.
}
\label{fig:resol final}
\end{center}
\end{figure}

Statistical fluctuations in $\theta_y^*$ are mostly due to the beam divergence and can be studied by comparing the angles reconstructed from the left and right arm. As illustrated in Figure~\ref{fig:beam div vert}, the distributions show only minimal deviations from a Gaussian shape. By dividing their standard deviation by a factor of 2, one can estimate the resolution of the two-arm reconstruction (Eq.~(\ref{eq:kin 2a})) of elastic events, see Figure~\ref{fig:resol final}, bottom. Moreover, measurements of beam emittances \cite{op-elog} indicate that the vertical divergences of the two beams can be considered as equal with a tolerance of about $25\un{\%}$. Exploiting this fact, one can de-convolute the distribution of $\theta_y^{*\rm R} - \theta_y^{*\rm L}$ in order to obtain the beam-divergence distribution, used for the acceptance corrections discussed in Section~\ref{sec:acc corr}.

In the horizontal projection, a more complex procedure is used since the one-arm reconstruction, Eq.~(\ref{eq:kin 1a}), is strongly influenced by the detector resolution. First, the horizontal beam divergence is estimated from the standard deviation of reconstructed vertices, $\sigma(x^*)$:
\begin{equation}
\label{eq:beam div from vertex}
\sigma^{\rm bd}(\theta_x^*) = {\sigma(x^*) \sqrt 2\over \beta^*}\ .
\end{equation}
It increases from $0.75$ to $0.9\un{\mu rad}$ over the time of the fill. Subtracting this component from the standard deviation of $\theta_x^{*\rm R} - \theta_x^{*\rm L}$, one determines the (mean) spatial resolution of the sensors in each diagonal: $10.7\un{\mu m}$ (45 top -- 56 bottom) and $12.1\un{\mu m}$ (45 bottom -- 56 top). These results have been verified to be time independent. Finally, the beam divergence and sensor resolution components can be propagated through Eq.~({\ref{eq:kin 2a}}) to estimate the $\theta_x^*$ resolution for elastic events, as plotted in Figure~\ref{fig:resol final}, top.

\subsection{Differential Cross-Section Reconstruction}
\label{sec:diff cs}

For a given $t$ bin, the differential cross-section is evaluated by selecting and counting elastic events:
\begin{equation}
{\d\sigma\over \d t}(\hbox{bin}) =
	{\cal N}\, {\cal U}({\rm bin})\, {\cal B}\, {1\over \Delta t}
	\sum\limits_{t\, \in\, {\rm bin}} {\cal A}(\theta^*, \theta_y^*)\ {\cal E}(\theta_y^*)
	\ ,
\end{equation}
where $\Delta t$ is the width of the bin, ${\cal N}$ is a normalisation factor and the other symbols stand for various correction factors:
 ${\cal U}$ for unfolding of resolution effects, ${\cal B}$ for background subtraction, ${\cal A}$ for acceptance correction and ${\cal E}$ for detection and reconstruction efficiency.


\subsubsection{Event Tagging}
\label{sec:tagging}

\begin{table}
\caption{The elastic selection cuts. The superscripts R and L refer to the right and left arm. The $\alpha \theta_x^*$ term in cut 3 absorbs the effects of residual optics imperfections, $\alpha$ is of the order of $0.1\un{\mu m/\mu rad}$. The right-most column gives a typical RMS of the cut distribution.
}
\label{tab:cuts}
\begin{center}
\begin{tabular}{ccc}\hline
number & cut & RMS ($\equiv 1\sigma$)\cr\hline
1 & $\theta_x^{*\rm R} - \theta_x^{*\rm L}$				& $3.9\un{\mu rad}$	\cr
2 & $\theta_y^{*\rm R} - \theta_y^{*\rm L}$				& $1.0\un{\mu rad}$	\cr
3 & $x^{*\rm R} - x^{*\rm L} - \alpha \theta_x^*$		& $250\un{\mu m}$ 	\cr\hline
\end{tabular}
\end{center}
\end{table}

\begin{figure}
\begin{center}
\includegraphics{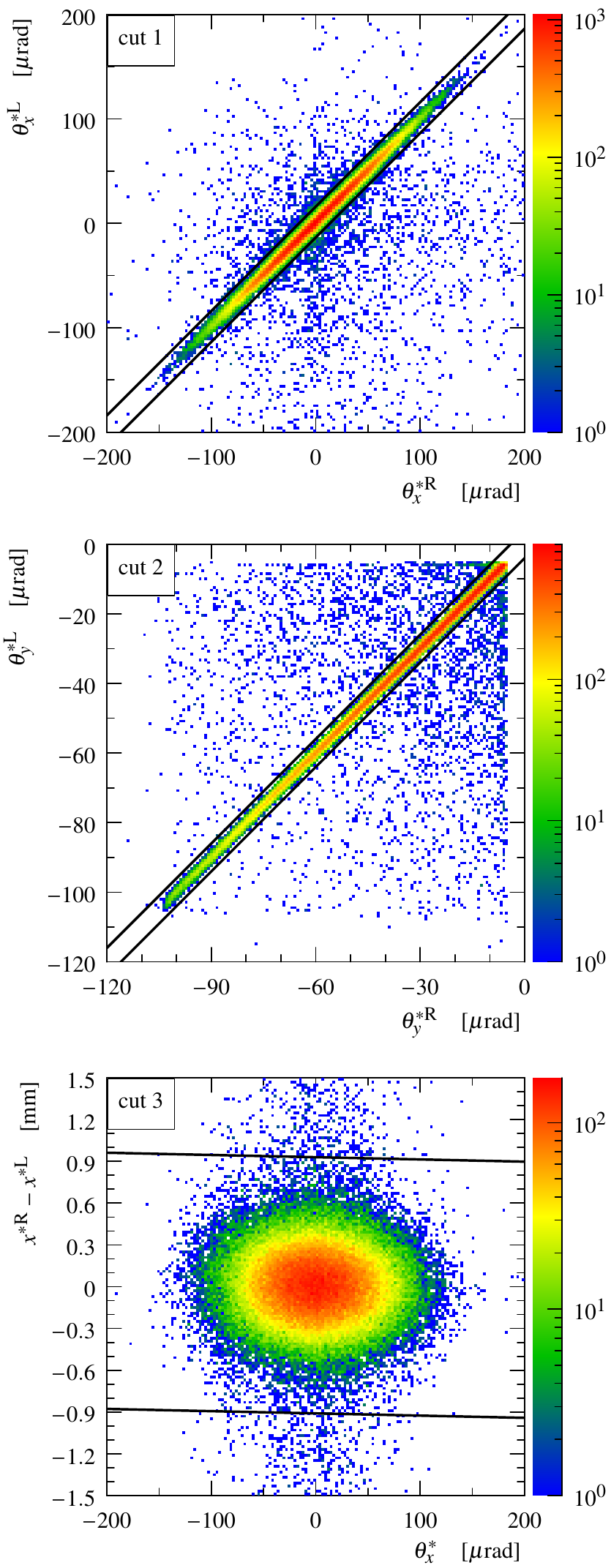}
\caption{%
Correlation plots for the event selection cuts presented in Table~\ref{tab:cuts}, showing events with diagonal topology 45 top -- 56 bottom. The solid black lines delimit the signal region within $\pm 4\un{\sigma}$.
}
\label{fig:cuts}
\end{center}
\end{figure}

The cuts used to select the elastic events are summarised in Table~\ref{tab:cuts}. Cuts 1 and 2 require the reconstructed-track\Break collinearity between the left and right arm. Cut 3 ensures that the protons come from the same vertex (horizontally). The correlation plots corresponding to these cuts are shown in Figure~\ref{fig:cuts}. Thanks to the very low beam divergence, the collinearity cuts are very powerful, and consequently other conceivable cuts (cf. Table~2 in~\cite{epl101-el}) bring no significant improvement.

Since a Monte-Carlo study shows that applying the three cuts at the $3\un{\sigma}$ level would lead to a loss of about $0.5\un{\%}$ of the elastic events, the cut threshold is set to $4\un{\sigma}$.

The tagging efficiency has been studied by applying the cuts also at the $5\un{\sigma}$-level. This selection has yielded $0.3\un{\%}$ more events in every $|t|$-bin. This kind of inefficiency only contributes to a global scale factor, which is irrelevant for this analysis because the normalisation is taken from a different data set (cf. Section~\ref{sec:normalisation}).


\subsubsection{Background}
\label{sec:background}

As the RPs were very close to the beam, one may expect an enhanced background from coincidence of beam halo protons hitting detectors in the two arms. Other background sources (pertinent to any elastic analysis) are: central diffraction and pile-up of two single diffraction events.

\begin{figure}
\begin{center}
\includegraphics{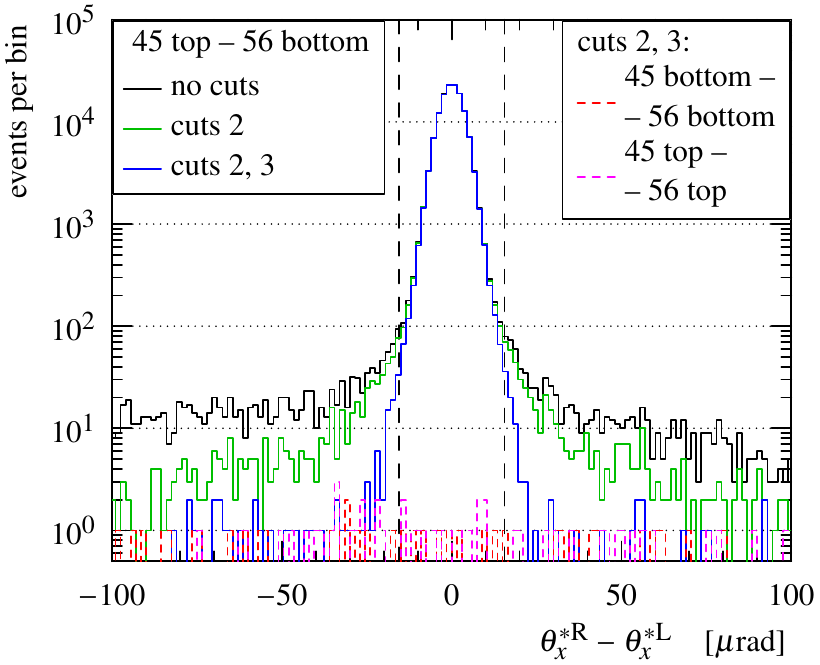}
\caption{%
Distributions of discriminator 1, i.e. the difference between the horizontal scattering angle reconstructed from the right and the left arm. Solid curves: data from diagonal 45 top -- 56 bottom, the different colours correspond to various combinations of the selection cuts (see numbering in Table~\ref{tab:cuts}). Dashed curves: data from anti-diagonal RP configurations, obtained by inverting track coordinates in the left arm. The vertical dashed lines represent the boundaries of the signal region ($\pm 4\un{\sigma}$).
}
\label{fig:tag bckg}
\end{center}
\end{figure}

The background rate (i.e.~impurity of the elastic tagging) is estimated in two steps, both based on distributions of discriminators from Table~\ref{tab:cuts} plotted in various situations, see an example in Figure~\ref{fig:tag bckg}. In the first step, diagonal data are studied under several cut combinations. While the central part (signal) remains essentially constant, the tails (background) are strongly suppressed when the number of cuts is increased. In the second step, the background distribution is interpolated from the tails into the signal region. The form of the interpolation is inferred from non-diagonal RP track configurations (\textit{45 bottom -- 56 bottom} or \textit{45 top -- 56 top}), artificially treated like diagonal signatures by inverting the coordinate signs in the arm 45; see the dashed distributions in the figure. These non-diagonal configurations cannot contain any elastic signal and hence consist purely of background which is expected to be similar in the diagonal and non-diagonal configurations. This expectation is supported by the agreement of the tails of the blue solid and dashed curves in the figure. Since the non-diagonal distributions are flat, the comparison of the signal-peak size to the amount of interpolated background yields the estimate $1 - {\cal B} < 10^{-4}$.

\subsubsection{Acceptance Correction}
\label{sec:acc corr}

\begin{figure}
\begin{center}
\includegraphics{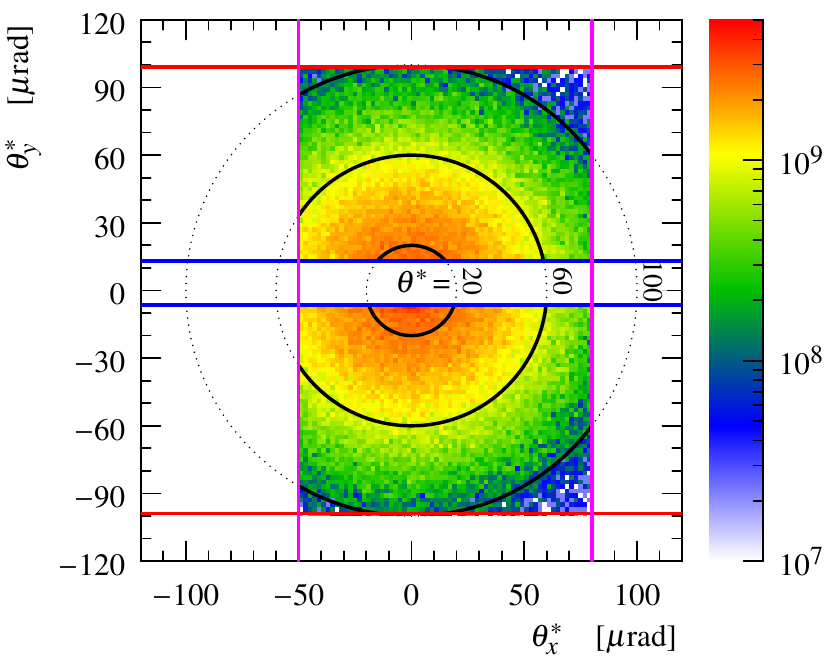}
\caption{%
Distribution of scattering angle projections $\theta_y^*$ vs.~$\theta_x^*$. The upper (lower) part comes from the diagonal 45 bottom -- 56 top (45 top -- 56 bottom). The red horizontal lines represent cuts due to the LHC apertures, the blue horizontal lines cuts due to the sensor edges. The vertical magenta lines delimit the fiducial region with detection efficiency not affected by the horizontal RPs. The dotted circles show contours of constant scattering angle $\theta^*$ as indicated in the middle of the plot (values in micro-radians). The parts of the contours within acceptance are emphasized in thick black.
}
\label{fig:acc corr princ}
\end{center}
\end{figure}

The acceptance of elastic protons is limited by two factors: sensor coverage (relevant for low $|\theta^*_y|$) and LHC beam aperture (at $|\theta^*_y| \approx 100\un{\mu rad}$). Moreover, there is a region in the kinematic parameter space where elastic protons may interact with the horizontal RPs leading to uncertain detection efficiency. To avoid this region, an additional fiducial cut has been adopted: $-50 < \theta_x^* < 80\un{\mu rad}$. In the far vertical RPs, this restriction corresponds to about $-2.3 < x < 3.7\un{mm}$. All acceptance related cuts are visualised in Figure~\ref{fig:acc corr princ}.

The correction for the above limitations includes two contributions -- a geometrical correction ${\cal A}_{\rm geom}$ reflecting the fraction of the phase space within the acceptance and a component ${\cal A}_{\rm fluct}$ correcting for fluctuations around the vertical acceptance limitations:
\begin{equation}
{\cal A}(\theta^*, \theta_y^*) = {\cal A}_{\rm geom}(\theta^*)\ {\cal A}_{\rm fluct}(\theta_y^*)\ .
\end{equation}
The fiducial cuts in $\theta_x^*$ have been given sufficient margin from the region with uncertain efficiency to render the respective fluctuation correction negligible.

The calculation of the geometrical correction ${\cal A}_{\rm geom}$ is based on the azimuthal symmetry of elastic scattering, experimentally verified for the data within acceptance. As\Break shown in Figure \ref{fig:acc corr princ}, for a given value of $\theta^*$ the correction is given by:
\begin{equation}
\label{eq:acc geom}
{\cal A_{\rm geom}}(\theta^*) = {
	\hbox{full circumference}\over 
	\hbox{arc length within acceptance}
} \ .
\end{equation}

The correction ${\cal A}_{\rm fluct}$ is calculated analytically from the probability that any of the two elastic protons leaves the region of acceptance due to the vertical beam divergence. The beam divergence distribution is modelled as a Gaussian with the spread determined by the method described in Section~\ref{sec:resolution}. This contribution is sizeable only close to the acceptance limitations. Data from regions with corrections larger than $2.5$ are discarded. The uncertainties are related to the resolution parameters. For the lowest $|t|$ bin their relative values are: vertical beam divergence: $2\un{\%}$, left-right asymmetry: $1\un{\%}$, and non-Gaussian shape: $1\un{\%}$.

\begin{figure}
\begin{center}
\includegraphics{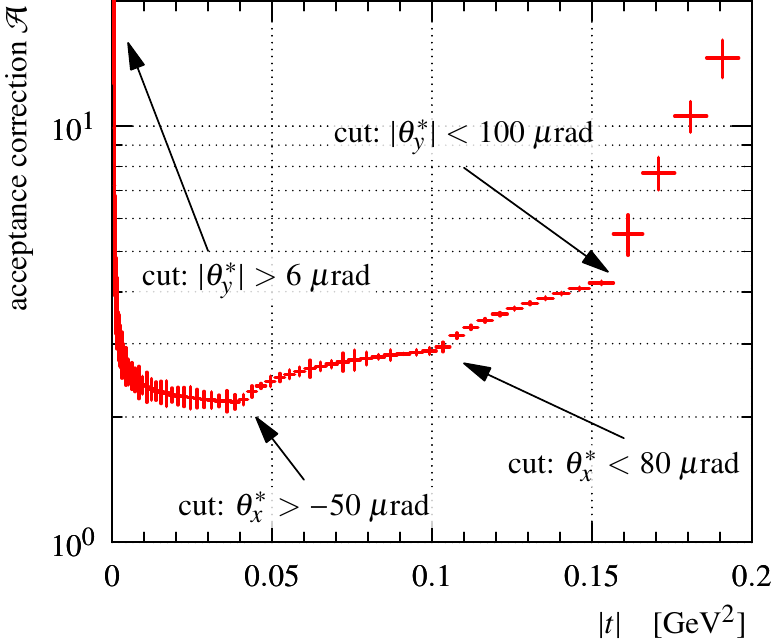}
\caption{%
Full acceptance correction, ${\cal A}$, for diagonal 45 bottom -- 56 top. The points give the mean value per bin, the error bars indicate the standard deviation. The abrupt changes in the shape correspond to acceptance cuts as indicated by the arrows.
}
\label{fig:acc corr res}
\end{center}
\end{figure}

Figure~\ref{fig:acc corr res} shows an example of the $t$-dependence of the acceptance correction for the diagonal reaching lower $|t|$-values. Since a single diagonal cannot cover more than half of the phase space, the minimum value of the correction is $2$. The very low $|t|$ data points with the full correction larger than $10$ are discarded to avoid biases. At the high-$|t|$ end all data points are kept.

\subsubsection{Inefficiency Corrections}
\label{sec:ineff corr}

Since the overall normalisation will be determined from another dataset (see Section~\ref{sec:normalisation}), any inefficiency correction that does not alter the $t$-distribution shape does not need to be considered in this analysis (trigger, data acquisition and pile-up inefficiency discussed in~\cite{epl101-el,prl111}). The remaining inefficiencies are related to the inability of a RP to resolve the elastic proton track.

\begin{figure}
\begin{center}
\includegraphics{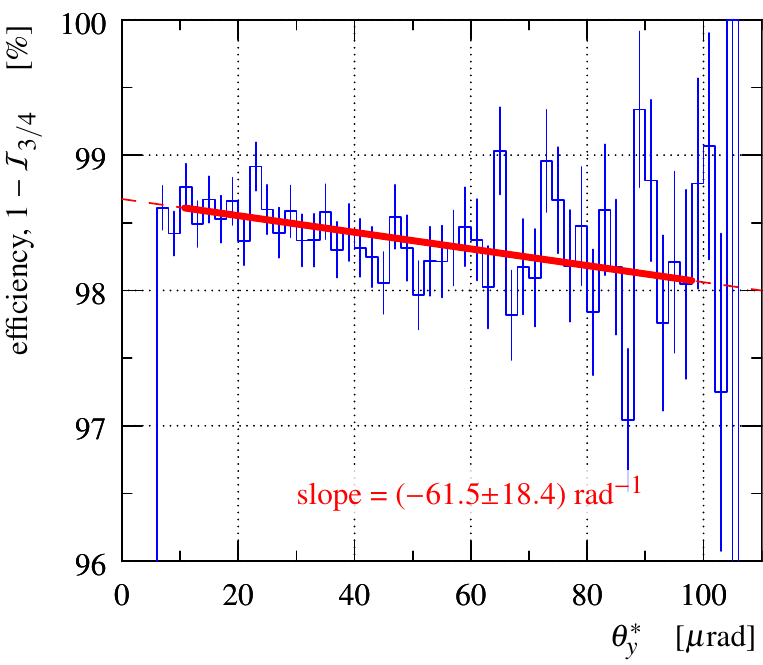}
\caption{%
Single-RP uncorrelated inefficiency for the near bottom RP in the right arm. The rapid drop at $\theta_y^* \approx 8\un{\mu rad}$ is due to acceptance effects at the sensor edge. The red lines represent a linear fit of the efficiency dependence on the vertical scattering angle (solid) and its extrapolation to the regions affected by acceptance effects (dashed).
}
\label{fig:eff 3/4}
\end{center}
\end{figure}

One such case is when a single RP does not detect and/or reconstruct a proton track, with no correlation to other RPs. This type of inefficiency, ${\cal I}_{3/4}$, is evaluated by removing the RP from the tagging cuts (Table \ref{tab:cuts}), repeating the event selection and calculating the fraction of recovered events. A typical example is given in Figure~\ref{fig:eff 3/4}, showing that the efficiency decreases gently with the vertical scattering angle. This dependence originates from the fact that protons with larger $|\theta_y^*|$ hit the RPs further from their edge and therefore the potentially created secondary particles have more chance to induce additional signal. Since the RP detectors cannot resolve multiple tracks (non-unique association between ``U'' and ``V'' track candidates), a secondary particle track prevents from using the affected RP in the analysis.

Another source of inefficiency are proton interactions in a near RP affecting simultaneously the far RP downstream. The contribution from these near-far correlated inefficiencies, ${\cal I}_{2/4}$, is determined by evaluating the rate of events with high track multiplicity ($\gtrsim$ 5) in both near and far RPs. Events with high track multiplicity simultaneously in a near top and near bottom RP are discarded as such a shower is likely to have started upstream from the RP station and thus be unrelated to the elastic proton interacting with detectors. The outcome, ${\cal I}_{2/4} \approx 1.5\un{\%}$, is compatible between left/right arms and top/bottom RP pairs and compares well to Monte-Carlo simulations (e.g.~section 7.5 in \cite{hubert-thesis}).

The full correction is calculated as
\begin{equation}
\label{efficiency}
	{\cal E}(\theta_y^*) = {1\over 1 - \left( \sum\limits_{i\in \rm RPs} {\cal I}^i_{3/4}(\theta_y^*) + 2 {\cal I}_{2/4} \right) } \ .
\end{equation}
The first term in the parentheses sums the contributions from the four RPs of a diagonal and increases from about $16$ to $18\un{\%}$ from the lowest to the highest $|\theta_y^*|$. These values are higher than in the previous analyses (e.g.~Section 5.2.4 in \cite{8tev-90m}) due the contribution from the far RPs in the left arm. The reconstruction efficiency in these pots is decreased by showers initiated by beam halo protons in the horizontal RPs upstream (closer to the beam in the left arm than in the right one).

\subsubsection{Unfolding of Resolution Effects}
\label{sec:unfolding}

Due to the very small beam divergence, the correction for resolution effects can be safely determined by the following iterative procedure.
\begin{itemize}
\item[1.] The differential cross-section data are fitted by a smooth curve.
\item[2.] The fit is used in a numerical-integration calculation of the smeared $t$-distribution (using the resolution parameters determined in Section~\ref{sec:resolution}). The ratio between the smeared and the non-smeared $t$-distributions gives a set of per-bin correction factors.
\item[3.] The corrections are applied to the observed (yet uncorrected) differential cross-section yielding a better estimate of the true $t$-distribution.
\item[4.] The corrected differential cross-section is fed back to step 1.
\end{itemize}
As the estimate of the true $t$-distribution improves, the difference between the correction factors obtained in two successive iterations decreases. When the difference becomes negligible, the iteration stops. This is typically achieved after the second iteration. 

\begin{figure}
\begin{center}
\includegraphics{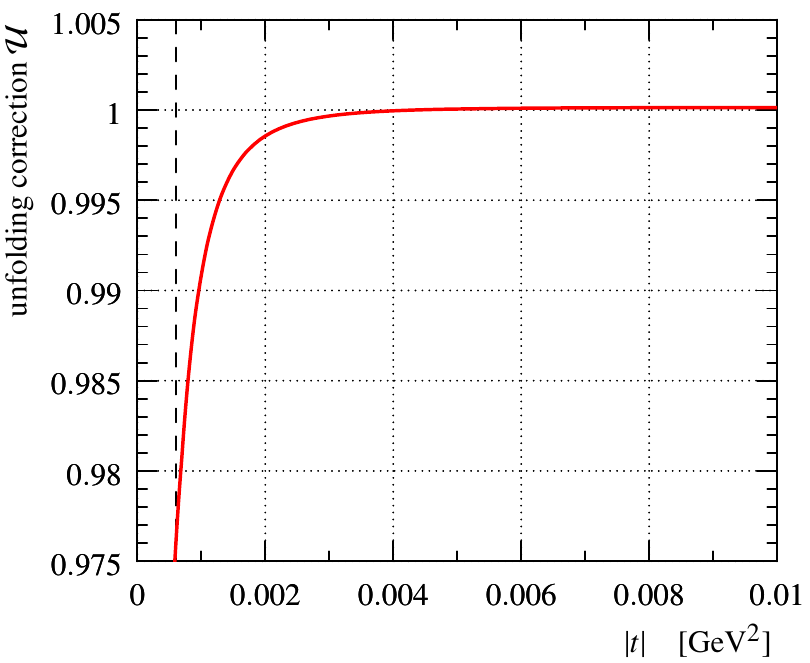}
\caption{%
Unfolding correction for the close diagonal (45 bottom -- 56 top). The vertical dashed line indicates the position of the acceptance cut due to sensor edges.
}
\label{fig:unfolding}
\end{center}
\end{figure}

The final correction is negligible (${\cal U} \approx 1$) for all bins except at very low $|t|$ where the rapid cross-section growth occurs, see Figure~\ref{fig:unfolding}.

For the uncertainty estimate, the uncertainties of the $\theta_x^*$ and $\theta_y^*$ resolutions (accommodating the full time variation) as well as fit-model dependence have been considered, each contribution giving a few per-mille for the lowest-$|t|$ bin.


\subsubsection{Normalisation}
\label{sec:normalisation}

The normalisation ${\cal N}$ is determined by requiring the same cross-section integral between $|t| = 0.014$ and $0.203\un{GeV^2}$ as for dataset 1 published in \cite{prl111}. This publication describes a measurement of elastic and inelastic rates at the same collision energy of $\sqrt s = 8\un{TeV}$. These rates can be combined using the optical theorem in order to resolve the value of the luminosity which consequently allows for normalisation of the differential cross-section. The leading uncertainty of ${\cal N}$, $4.2\un{\%}$, comes from the rate uncertainties in \cite{prl111}.


\subsubsection{Binning}
\label{sec:binning}

At very low $|t|$, where the cross-section varies the fastest ($\approx 0.001\un{GeV^2}$), a fine binning is used. In the middle of the $|t|$ range ($\approx 0.03\un{GeV^2}$), the bin width is chosen to give about $1\un{\%}$ statistical uncertainty. This rule is abandoned at higher $|t|$ (above $0.07\un{GeV^2}$) in favour of bins with a constant width of $0.01\un{GeV^2}$ to avoid excessively large bins.


\subsubsection{Systematic Uncertainties}
\label{sec:systematics}

Besides the systematic uncertainties mentioned at the above analysis steps, the beam momentum uncertainty needs to be considered when the scattering angles are translated to $t$, see Eq.~(\ref{eq:th t}). The uncertainty was estimated to $0.1\un{\%}$ in Section 5.2.8 in \cite{8tev-90m} which is further supported by a recent review \cite{todesco-lpc}.

Two different methods are used to propagate the systematic effects into the $t$-distribution. The first is based on a Monte-Carlo simulation which uses a fit of the final differential cross-section data to generate the true $t$-distribution. In parallel, another $t$-distribution is built, introducing one of the above mentioned systematic effects at $1\un{\sigma}$ level. The difference between the two $t$-distributions gives the systematic effect on the differential cross-section. The second method is similar, however using numerical integration techniques instead of Monte-Carlo simulations. Both methods are formally equivalent to evaluating
\begin{equation}
\label{eq:syst mode}
\delta s_{q}(t) \equiv \frac{\partial(\d\sigma/\d t)}{\partial q}\ \delta q\ ,
\end{equation}
where $\delta q$ corresponds to a $1\un{\sigma}$ bias in the quantity $q$ responsible for a given systematic effect.

The Monte-Carlo simulations show that the combined effect of several systematic errors is well approximated by linear combination of the individual contributions from\Break Eq.~(\ref{eq:syst mode}).


\subsection{Systematic Cross-Checks}
\label{sec:cross checks}

Compatible results have been obtained by analysing data subsets of events from different bunches, different diagonals and different time periods -- in particular those right after and right before the beam cleanings.

\subsection{Final Data Merging}
\label{sec:final data merging}

Finally, the differential cross-section histograms from both diagonals are merged. This is accomplished by a per-bin weighted average, with the weight given by inverse squared statistical uncertainty. The statistical and systematic uncertainties are propagated accordingly. For the systematic ones, the correlation between the diagonals is taken into account. For example the vertical (mis-)alignment of the RPs of one unit is almost fully correlated; thus the effect on the differential cross-section is opposite for the two diagonals and consequently its impact is strongly reduced once the diagonals are merged.

\begin{figure*}
\vskip-5mm
\begin{center}
\includegraphics{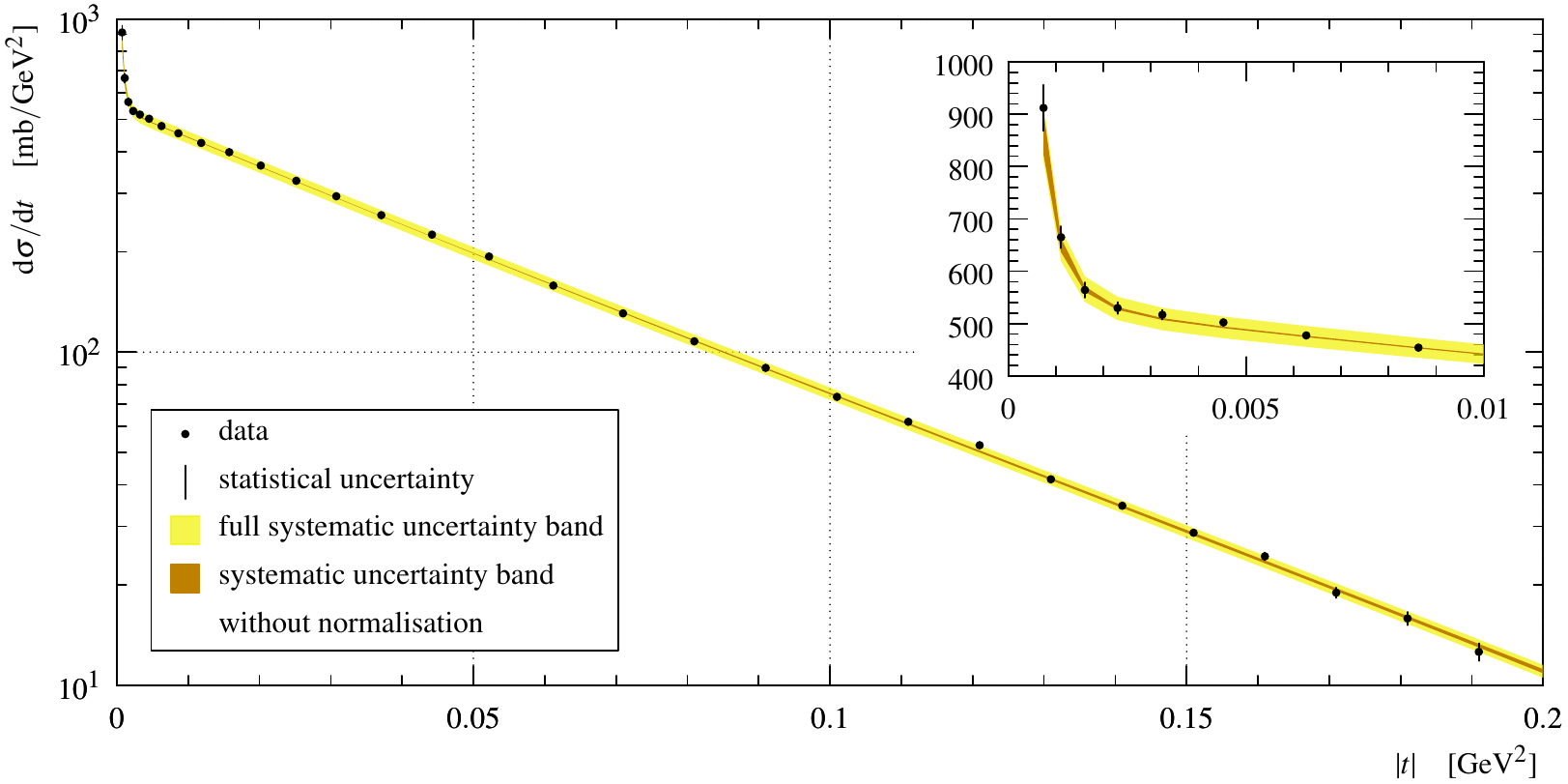}
\vskip-3mm
\caption{%
Differential cross-section from Table \ref{tab:data} with statistical (bars) and systematic uncertainties (bands). The grey band represents all systematic uncertainties, the brown one all but normalisation. The bands are centred around a data fit including both nuclear and Coulomb components (Eqs.~(\ref{eq:int kl}), (\ref{eq:nuc phase con}) and (\ref{eq:nuc mod}) with $N_b=3$). INSET: a low-$|t|$ zoom featuring cross-section rise due to the Coulomb interaction.
}
\label{fig:dsdt}
\end{center}
\end{figure*}

The cross-section values can be found in Table~\ref{tab:data} and visualised in Figure~\ref{fig:dsdt}. The figure clearly shows a rapid cross-section rise below $|t| \lesssim 0.002\un{GeV^2}$, which will later be interpreted as an effect due to electromagnetic interaction.

\input data_table.tex

\begin{figure*}
\begin{center}
\includegraphics{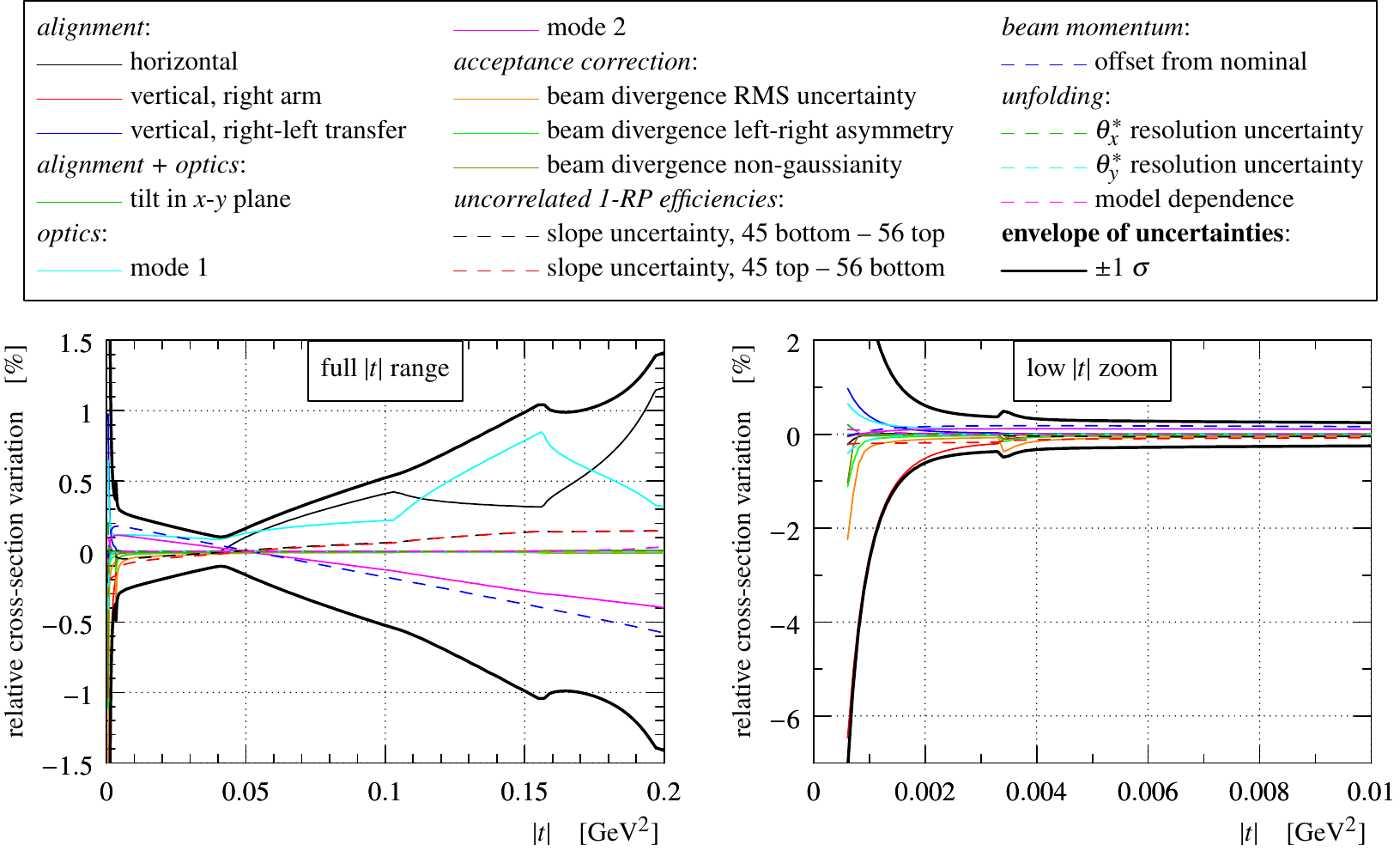}
\caption{%
Impact of $t$-dependent systematic effects on the differential cross-section. Each curve corresponds to a systematic error of $1\un{\sigma}$, cf.~Eq.~(\ref{eq:syst mode}). The two contributions due to the optics correspond to the two vectors in Eq.~(\ref{eq:opt bias modes}). The envelope is determined by summing all shown contributions in quadrature for each $|t|$ value.
}
\label{fig:syst unc}
\end{center}
\end{figure*}

The final systematic uncertainties, except the $4.2\un{\%}$ coming from the normalisation, are summarised in Figure~\ref{fig:syst unc} where their impact on the differential cross-section is shown. The leading uncertainties include normalisation, optics imperfections, beam momentum offset and residual misalignment. The vertical misalignment is the dominant systematic effect in the very-low $|t|$ region. The leading effects are quantified in Table~\ref{tab:data} and can be used to approximate the covariance matrix of systematic uncertainties:
\begin{equation}
\label{eq:covar mat}
\mat V_{ij} = \sum_{q} \delta s_{q}(i)\ \delta s_{q}(j)\: ,
\end{equation}
where $i$ and $j$ are bin indices (row numbers in Table~\ref{tab:data}) and the sum goes over the leading error contributions $q$ (six right-most columns in the table).

Let us emphasize that the systematic effects with linear $t$ dependence (see Figure~\ref{fig:syst unc}) cannot alter the non-purely-exponential character of the data. This is the case for the effects of normalisation, beam momentum and to a large degree also of optics-mode 2. For the beam momentum, this can also be understood analytically: changing the value of $p$ would yield a scaling of $t$, see Eq.~(\ref{eq:th t}), and consequently also scaling of the $b_n$ parameters in Eq.~(\ref{eq:nuc mod}). However, the non-zero $b_2, b_3$ etc.~parameters (reflecting the non-exponentiality) cannot be brought to $0$ (as in purely-exponential case).

%% file: data_table.tex
\def\tableHeader{%
	\multispan3\vrule width0pt height9pt depth3pt\hss $|t|$ bin $\unt{GeV^2}$\hss & \multispan9\hss $\d\sigma/\d t \ung{mb/GeV^2}$ \hss \cr
	\multispan3\hrulefill\hbox to12pt{\hfil} & \multispan9\hrulefill\cr
	left & right & represent. & value & statist.     & system.  & normal.    & optics   & optics   & beam		& alignment	& alignment\cr
	edge & edge  & point      &       & uncert.      & uncert.  & ${\cal N}$ & mode 1   & mode 2   & momentum	& hor.~shift	& vert.~shift\cr
}

\begin{table*}
\caption{%
The elastic differential cross-section as determined in this analysis. The three left-most columns describe the bins in $t$. The representative point gives the $t$ value suitable for fitting~\cite{lafferty94}.
The other columns are related to the differential cross-section. The six right-most columns give the leading systematic biases in $\d\sigma/\d t$ for $1\sigma$-shifts in the respective quantities, $\delta s_q$, see Eqs.~(\ref{eq:syst mode}) and (\ref{eq:covar mat}). The two contributions due to optics correspond to the two vectors in Eq.~(\ref{eq:opt bias modes}).
}%
\label{tab:data}
\begin{center}
\setlength{\tabcolsep}{5pt}
\def\arraystretch{0.8}
\begin{tabular}{ccc@{\hskip15pt}ccccccccc}
\hline
\tableHeader
\hline
$0.000600$ & $0.000916$ & $0.000741$ & $912.13$ & $44.0\S$ & $54.7\S$ & $+36.3\S$ & $+4.18\S$ & $-0.151$ & $+0.029$ & $-0.791$ & $-40.7\S\S$ \cr
$0.000916$ & $0.001346$ & $0.001110$ & $665.09$ & $21.0\S$ & $30.4\S$ & $+27.3\S$ & $+1.66\S$ & $+0.318$ & $+0.579$ & $+0.042$ & $-13.5\S\S$ \cr
$0.001346$ & $0.001930$ & $0.001612$ & $564.20$ & $14.6\S$ & $24.3\S$ & $+23.8\S$ & $+0.905$ & $+0.500$ & $+0.806$ & $+0.065$ & $-\S4.87\S$ \cr
$0.001930$ & $0.002725$ & $0.002298$ & $529.76$ & $11.3\S$ & $22.4\S$ & $+22.2\S$ & $+0.663$ & $+0.569$ & $+0.895$ & $+0.027$ & $-\S1.99\S$ \cr
$0.002725$ & $0.003806$ & $0.003240$ & $516.92$ & $\S9.19$ & $21.4\S$ & $+21.4\S$ & $+0.579$ & $+0.585$ & $+0.914$ & $+0.004$ & $-\S1.05\S$ \cr
$0.003806$ & $0.005276$ & $0.004525$ & $502.29$ & $\S6.24$ & $20.8\S$ & $+20.7\S$ & $+0.587$ & $+0.570$ & $+0.891$ & $-0.008$ & $-\S0.216$ \cr
$0.005276$ & $0.007276$ & $0.006266$ & $477.43$ & $\S4.83$ & $20.0\S$ & $+20.0\S$ & $+0.563$ & $+0.536$ & $+0.840$ & $-0.010$ & $-\S0.126$ \cr
$0.007276$ & $0.009995$ & $0.008628$ & $454.13$ & $\S3.86$ & $19.1\S$ & $+19.1\S$ & $+0.534$ & $+0.486$ & $+0.763$ & $-0.010$ & $-\S0.089$ \cr
$0.009995$ & $0.01369\S$ & $0.01183\S$ & $424.90$ & $\S3.09$ & $17.9\S$ & $+17.9\S$ & $+0.496$ & $+0.421$ & $+0.663$ & $-0.004$ & $-\S0.061$ \cr
$0.01369\S$ & $0.01786\S$ & $0.01576\S$ & $398.49$ & $\S2.75$ & $16.5\S$ & $+16.5\S$ & $+0.447$ & $+0.349$ & $+0.552$ & $-0.005$ & $-\S0.048$ \cr
$0.01786\S$ & $0.02255\S$ & $0.02019\S$ & $363.33$ & $\S2.44$ & $15.1\S$ & $+15.1\S$ & $+0.394$ & $+0.279$ & $+0.443$ & $-0.002$ & $-\S0.035$ \cr
$0.02255\S$ & $0.02783\S$ & $0.02517\S$ & $327.03$ & $\S2.15$ & $13.7\S$ & $+13.7\S$ & $+0.338$ & $+0.210$ & $+0.337$ & $-0.006$ & $-\S0.031$ \cr
$0.02783\S$ & $0.03378\S$ & $0.03077\S$ & $293.88$ & $\S1.90$ & $12.2\S$ & $+12.2\S$ & $+0.283$ & $+0.147$ & $+0.238$ & $-0.005$ & $-\S0.025$ \cr
$0.03378\S$ & $0.04047\S$ & $0.03709\S$ & $257.86$ & $\S1.67$ & $10.8\S$ & $+10.8\S$ & $+0.229$ & $+0.089$ & $+0.148$ & $-0.005$ & $-\S0.020$ \cr
$0.04047\S$ & $0.04801\S$ & $0.04419\S$ & $225.35$ & $\S1.49$ & $\S9.34$ & $+\S9.34$ & $+0.229$ & $+0.036$ & $+0.068$ & $+0.088$ & $-\S0.007$ \cr
$0.04801\S$ & $0.05650\S$ & $0.05220\S$ & $193.69$ & $\S1.35$ & $\S7.98$ & $+\S7.97$ & $+0.261$ & $-0.011$ & $-0.000$ & $+0.232$ & $-\S0.006$ \cr
$0.05650\S$ & $0.06606\S$ & $0.06121\S$ & $158.48$ & $\S1.18$ & $\S6.69$ & $+\S6.68$ & $+0.258$ & $-0.047$ & $-0.054$ & $+0.306$ & $-\S0.004$ \cr
$0.06606\S$ & $0.07606\S$ & $0.07098\S$ & $130.78$ & $\S1.06$ & $\S5.54$ & $+\S5.52$ & $+0.239$ & $-0.072$ & $-0.094$ & $+0.337$ & $-\S0.003$ \cr
$0.07606\S$ & $0.08606\S$ & $0.08098\S$ & $107.80$ & $\S0.98$ & $\S4.57$ & $+\S4.55$ & $+0.214$ & $-0.087$ & $-0.118$ & $+0.340$ & $-\S0.002$ \cr
$0.08606\S$ & $0.09606\S$ & $0.09098\S$ & $\S89.71$ & $\S0.90$ & $\S3.77$ & $+\S3.75$ & $+0.188$ & $-0.095$ & $-0.131$ & $+0.328$ & $-\S0.001$ \cr
$0.09606\S$ & $0.1061\S\S$ & $0.1010\S\S$ & $\S73.41$ & $\S0.83$ & $\S3.12$ & $+\S3.10$ & $+0.163$ & $-0.097$ & $-0.136$ & $+0.306$ & $-\S0.000$ \cr
$0.1061\S\S$ & $0.1161\S\S$ & $0.1110\S\S$ & $\S61.78$ & $\S0.79$ & $\S2.58$ & $+\S2.56$ & $+0.214$ & $-0.099$ & $-0.136$ & $+0.234$ & $+\S0.001$ \cr
$0.1161\S\S$ & $0.1261\S\S$ & $0.1210\S\S$ & $\S52.55$ & $\S0.76$ & $\S2.14$ & $+\S2.11$ & $+0.241$ & $-0.097$ & $-0.131$ & $+0.179$ & $+\S0.001$ \cr
$0.1261\S\S$ & $0.1361\S\S$ & $0.1310\S\S$ & $\S41.52$ & $\S0.70$ & $\S1.78$ & $+\S1.75$ & $+0.246$ & $-0.093$ & $-0.125$ & $+0.141$ & $+\S0.001$ \cr
$0.1361\S\S$ & $0.1461\S\S$ & $0.1410\S\S$ & $\S34.58$ & $\S0.66$ & $\S1.48$ & $+\S1.44$ & $+0.239$ & $-0.087$ & $-0.116$ & $+0.113$ & $+\S0.001$ \cr
$0.1461\S\S$ & $0.1561\S\S$ & $0.1510\S\S$ & $\S28.69$ & $\S0.61$ & $\S1.23$ & $+\S1.19$ & $+0.227$ & $-0.080$ & $-0.107$ & $+0.091$ & $+\S0.000$ \cr
$0.1561\S\S$ & $0.1661\S\S$ & $0.1610\S\S$ & $\S24.37$ & $\S0.65$ & $\S1.01$ & $+\S0.99$ & $+0.169$ & $-0.072$ & $-0.098$ & $+0.095$ & $+\S0.000$ \cr
$0.1661\S\S$ & $0.1761\S\S$ & $0.1710\S\S$ & $\S18.95$ & $\S0.68$ & $\S0.84$ & $+\S0.81$ & $+0.117$ & $-0.064$ & $-0.088$ & $+0.104$ & $+\S0.000$ \cr
$0.1761\S\S$ & $0.1861\S\S$ & $0.1810\S\S$ & $\S15.86$ & $\S0.73$ & $\S0.69$ & $+\S0.67$ & $+0.082$ & $-0.056$ & $-0.079$ & $+0.111$ & $+\S0.000$ \cr
$0.1861\S\S$ & $0.1961\S\S$ & $0.1910\S\S$ & $\S12.59$ & $\S0.77$ & $\S0.58$ & $+\S0.55$ & $+0.054$ & $-0.049$ & $-0.071$ & $+0.123$ & $+\S0.000$ \cr
\hline
\end{tabular}
\end{center}
\end{table*}

%% file: coulomb.tex
\section{Coulomb-Nuclear Interference}
\label{sec:coulomb}

The Coulomb-nuclear interference (CNI) can be used to\Break probe the nuclear component of the scattering amplitude. Since the CNI effects are sensitive to the phase of the nuclear amplitude, both modulus and phase can be tested. 

For the modulus, a relevant question is whether the earlier reported non-exponentiality of the differential\Break cross-section \cite{8tev-90m} can be attributed solely to the nuclear\Break component or whether Coulomb scattering gives a sizeable contribution. Concerning the phase, several parametrisations\Break with different physics interpretations will be tested; for each of them the $\rho$ parameter (representative for the phase value at $t = 0$ according to Eq.~(\ref{eq:rho def})) will be determined.

Section~\ref{sec:cni framework} outlines the theoretical concepts needed to describe the CNI effects. Section~\ref{sec:cni anal proc} provides details on fitting procedures used to analyse the data. Sections \ref{sec:fit exp1} and \ref{sec:fit exp3} discuss the fit results for two relevant alternatives in the description of the nuclear modulus: either exponential functions with exponents linear in $t$ (called ``purely exponential'') or exponential functions with higher-degree polynomials of $t$ in the exponent (called ``non-exponential'').


\subsection{Theoretical Framework}
\label{sec:cni framework}

The amplitude describing elastic scattering of protons may be expected to receive three contributions, each corresponding to one of the following sets of Feynman diagrams.
\begin{itemize}
\item Containing QED elements only. This amplitude can be obtained by perturbative calculations, see Section~\ref{sec:cni coulomb}.
\item Containing QCD elements only. This amplitude is not directly calculable from the QCD lagrangian,\Break Sections~\ref{sec:cni nuclear modulus} and \ref{sec:cni nuclear phase} will propose several phenomenologically motived parame\-trisations.
\item Containing both QED and QCD elements. This contribution can neither be directly calculated from the Lagrangians, nor can \textit{ad hoc} parametrisations be used -- this amplitude is correlated with the previous two. Section~\ref{sec:cni interference} will introduce several interference formulae attempting to calculate the corresponding effects.
\end{itemize}

\subsubsection{Coulomb Amplitude}
\label{sec:cni coulomb}
The Coulomb amplitude can be calculated from QED\Break (e.g.~\cite{block96}), using empirical electric ${\cal F}_{\rm E}$ and magnetic ${\cal F}_{\rm M}$ form factors of the proton. It can be shown (e.g.~Section 1.3.1 in~\cite{jan_thesis}) that, at low $|t|$, the effect of both form factors can be described by a single function ${\cal F}$:
\begin{equation}
\label{eq:coul cs}
	{\d\sigma^{\rm C}\over \d t} = {4\pi\alpha^2\over t^2}\,{\cal F}^4\ ,\ 
	{\cal F}^2 = {{\cal F}_{\rm E}^2 + \tau {\cal F}_{\rm M}^2\over 1 + \tau}\ ,\ 
	\tau = {|t|\over 4m^2}\ ,
\end{equation}
where $\alpha$ is the fine-structure constant and $m$ represents the proton mass.

\subsubsection{Nuclear Amplitude -- Modulus}
\label{sec:cni nuclear modulus}

At $|t| \gtrsim 0.02\un{GeV^2}$ the effects due to the Coulomb interaction are not expected to be large (c.f.~Figure~\ref{fig:cni effect} or \cite{kklp11}). Thus, the measured cross-section can be attributed -- to a large extent -- to the nuclear component. Following Table~\ref{tab:data} and our previous publication \cite{8tev-90m} with high-precision data for $|t| < 0.2\un{GeV^2}$, the nuclear modulus will be parametrised as
\begin{equation}
\label{eq:nuc mod}
\left | {\cal A}^{\rm N}(t) \right | = \sqrt{s\over\pi} {p\over \hbar c} \sqrt{a} \exp\left( {1\over 2} \sum\limits_{n = 1}^{N_b} b_n\, t^n \right)\ ,
\end{equation}
where $N_b$ is the number of free parameters in the exponent. Consistently with \cite{8tev-90m} \footnote{%
Please note that Eq.~(15) in \cite{8tev-90m} contains a misprint: the exponent should have read $\sum\limits_{i=1}^{N_b} b_i\, |t|^i$.
}, the parameter $b_1$ gives the forward diffractive slope and $a$ the intercept of the differential cross-section at $t=0$. This parametrisation is also compatible with a number of theoretical models (see e.g.~\cite{elegent}).

Since the calculation of CNI may, in principle, involve integrations (e.g.~Eq.~(\ref{eq:int kl})), it is necessary to extend the nuclear amplitude meaningfully to $|t| > 0.2\un{GeV^2}$. Therefore the parametrisation Eq.~(\ref{eq:nuc mod}) is only used for $|t| < 0.2\un{GeV^2}$ while at $|t| > 0.5\un{GeV^2}$ the amplitude is fixed to follow a preliminary cross-section derived from the same data set as in \cite{8tev-90m} which features a dip-bump structure similar to the one observed at $\sqrt{s} = 7\un{TeV}$ \cite{epl95}. In order to avoid numerical problems, the intermediate region $0.2 < |t| < 0.5\un{GeV^2}$ is modelled with a continuous and smooth interpolation between the low and high-$|t|$ parts. It will be shown that altering the extended part of the nuclear amplitude\Break ($|t| > 0.2\un{GeV^2}$) within reasonable limits has negligible impact on the results presented later on.

\subsubsection{Nuclear Amplitude -- Phase}
\label{sec:cni nuclear phase}

The following phase parametrisations are considered.

\begin{figure*}
\begin{center}
\includegraphics{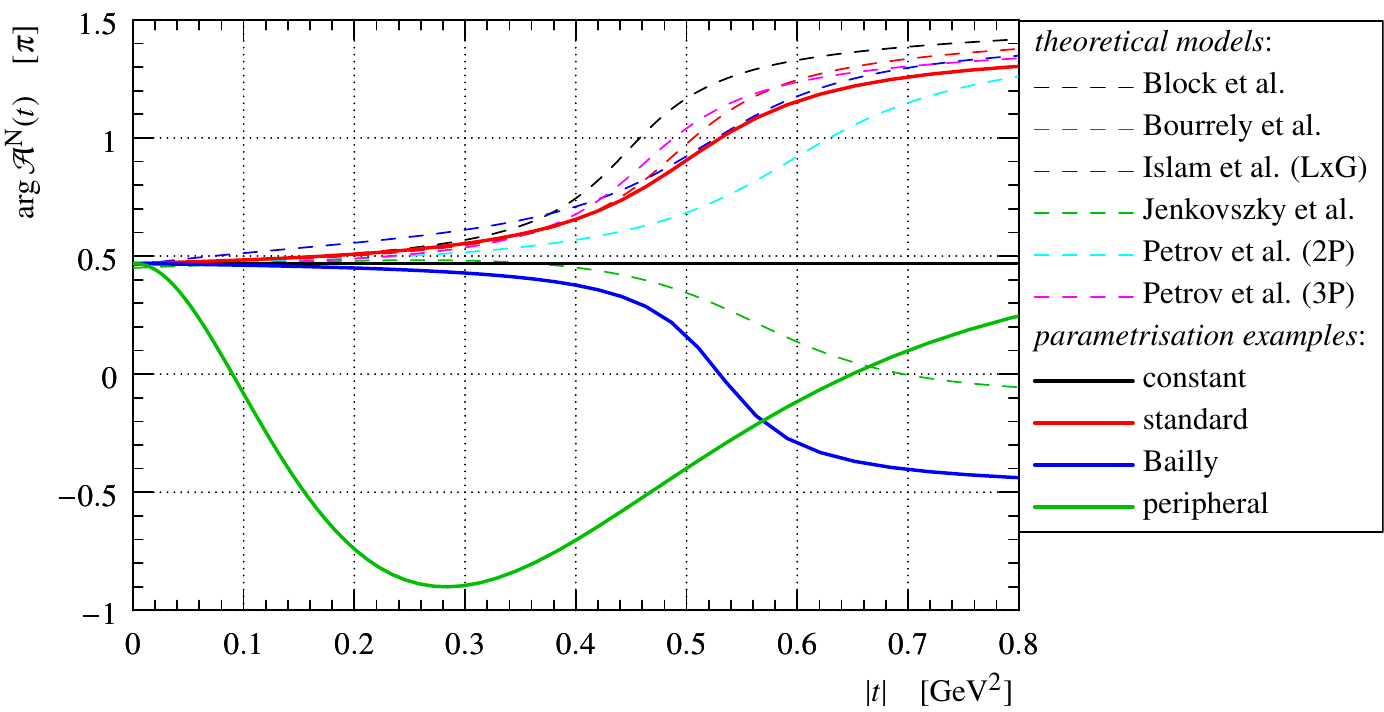}
\caption{Illustration of nuclear-phase forms. The dashed lines correspond to predictions by theoretical models (\cite{elegent} and references therein). The solid lines are typical examples of phases used in this report, all with the same value of $\rho = 0.10$. The peripheral example corresponds to the parameter values in Eq.~(\ref{eq:nuc phase per val}).
}
\label{fig:phase illustration}
\end{center}
\end{figure*}

\begin{enumerate}

\item[a)]
A {\bf constant phase} is obviously the simplest choice:
\begin{equation}
\label{eq:nuc phase con}
\arg {\cal A}^{\rm N}(t) = {\pi\over 2} - \arctan\rho = \hbox{const} \ .
\end{equation}
It leads to a strict proportionality between the real and the imaginary part of the amplitude at all $t$.

\item[b)]
The {\bf standard phase} parametrisation,
\begin{equation}
\label{eqn:nuc phase std}
	\begin{aligned}
		\arg {\cal A}^{\rm N}(t) =	& {\pi\over 2} - \arctan\rho + \arctan \left(\frac{|t|-|t_{0}|}{\tau}\right)\cr
									&- \arctan \left(\frac{-|t_{0}|}{\tau}\right) \: ,
	\end{aligned}
\end{equation}
describes the main features of many theoretical models -- almost imaginary amplitude in the forward direction ($\rho$ small) while almost purely real at the diffraction dip. The parameter values $t_0 = - 0.50\un{GeV^2}$ and $\tau = 0.1\un{GeV^2}$ have been chosen such that the shape is similar to a number of model predictions, see Figure~\ref{fig:phase illustration}.

\item[c)]
The parametrisation by {\bf Bailly et al.}~\cite{bailly87}:
\begin{equation}
\label{eq:nuc phase bai}
	\arg {\cal A}^{\rm N}(t) = {\pi\over 2} - \arctan {\rho\over 1 - {t\over t_{\rm d}}}
\end{equation}
where $t_{\rm d} \approx -0.53\un{GeV^2}$ gives the position of the diffractive minimum at $8\un{TeV}$ (preliminary result derived from the $\beta^* = 90\un{m}$ data \cite{8tev-90m}). This phase has a behaviour qualitatively similar to the model of Jenkovszky et al., see Figure~\ref{fig:phase illustration}.

\item[d)]
Another parametrisation was proposed in \cite{kl94}:
\begin{equation}
\label{eq:nuc phase per}
\arg {\cal A}^{\rm N}(t) = {\pi\over 2} - \arctan\rho - \zeta_1 \left(- {t\over 1\un{GeV^2}} \right)^\kappa \e^{\nu t}\ .
\end{equation}
As shown in Figure~\ref{fig:phase illustration}, it features a peak at $t = -\kappa / \nu$ and for asymptotically increasing $|t|$ it returns to its value at $t=0$. Due to a potentially rapid variation at low $|t|$, this functional form can yield an impact-parameter-space behaviour that is qualitatively different from the one obtained with the above parametrisations. In order to ensure fit stability, the parameters
\begin{equation}
\label{eq:nuc phase per val}
	\zeta_1 = 800\ ,\quad
	\kappa = 2.311\ ,\quad
	\nu = 8.161\un{GeV^{-2}}
\end{equation}
have been fixed to example values maintaining the desired impact-parameter behaviour at $\sqrt s = 8\un{TeV}$, using a method detailed in \cite{pk16}. This parametrisation with one free parameter will be denoted as {\bf peripheral phase} in what follows.

\end{enumerate}

Figure~\ref{fig:phase illustration} shows on the same plot a comparison of phase predictions by several models to typical examples of\Break parametrisations proposed above.

It should be noted that the nuclear phase has a strong influence on the amplitude behaviour in the space of impact parameter $b$ (for a detailed discussion see e.g.~Section~3 in~\cite{klk02}). A particularly decisive feature is the rate of phase variation at low $|t|$. Looking at Figure~\ref{fig:phase illustration} one can see that the constant, standard and Bailly phases are essentially flat at low $|t|$, thus leading to qualitatively similar pictures in the impact parameter space: elastic collisions being more central (preferring lower values of $b$) than the inelastic ones. Conversely, the peripheral phase parametrisation can yield a description with the opposite hierarchy, which is argued to be more natural by some authors (e.g. Section~4 in~\cite{kl96}). An impact-parameter study of the presented data will be given at end of Section~\ref{sec:fit exp3}.

\subsubsection{Coulomb-Nuclear Interference Formulae}
\label{sec:cni interference}

\begin{figure*}
\begin{center}
\includegraphics{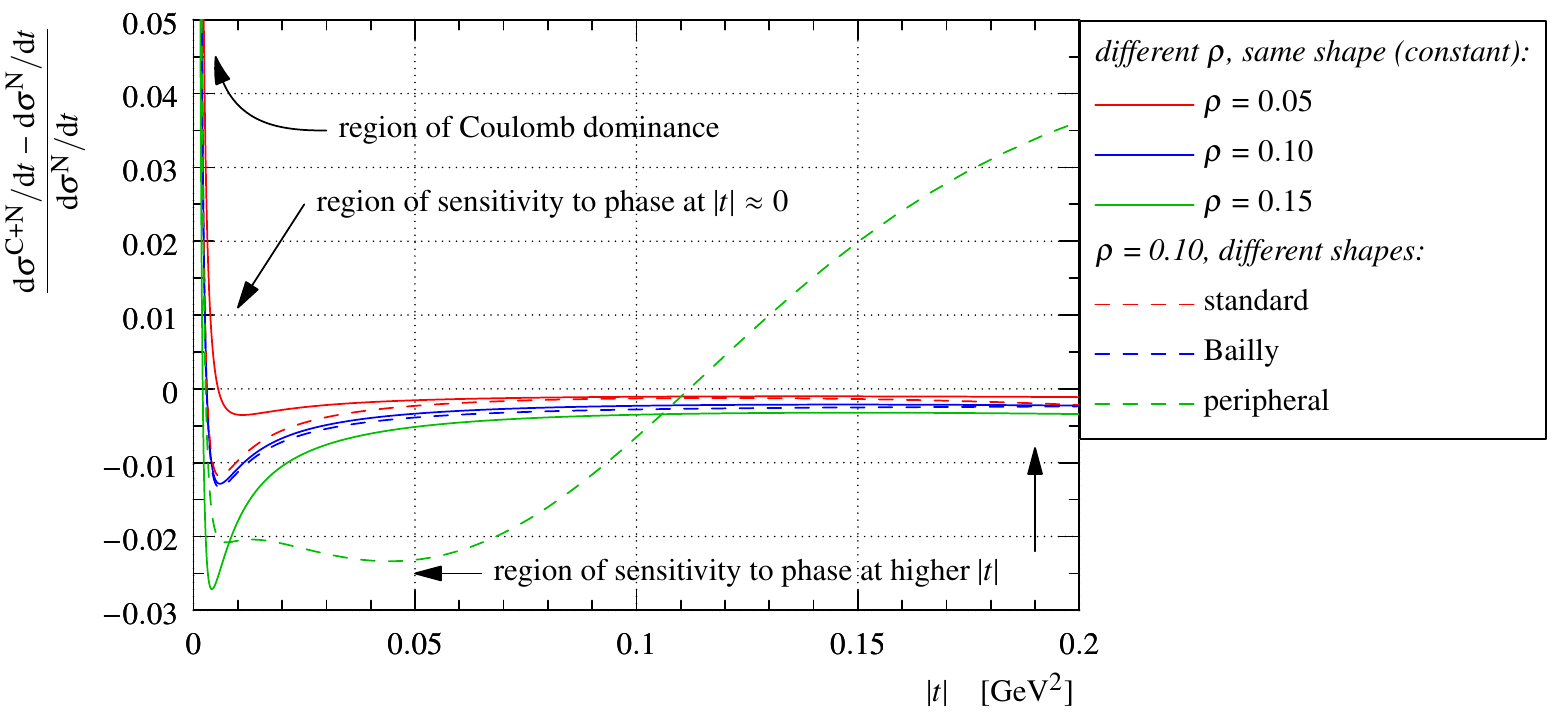}
\caption{%
Illustration of the effects due to the Coulomb interaction, using the KL formula. With the Cahn formula the plot looks identical. For the SWY formula, the picture is similar, however it misses the effects at $|t| \gtrsim 0.02\un{GeV^2}$. The curves show a response of the interference formula to different nuclear phases with a purely exponential nuclear modulus. The solid curves correspond to phases of the same shape (constant) but different values of $\rho$: the maximal response can be seen at $|t| \lesssim 0.01\un{GeV^2}$. Conversely, the dashed lines correspond to phases with fixed $\rho$ but various shapes (the same examples as in Figure~\ref{fig:phase illustration}): the response may be sizeable (e.g.~in the peripheral case) also at $|t| \gtrsim 0.02\un{GeV^2}$.
}
\label{fig:cni effect}
\end{center}
\end{figure*}

The {\bf simplified West-Yennie formula (SWY)} \cite{wy68} was derived in the framework of perturbative quantum field theory by evaluating the lowest-order Feynman diagrams that comprise both nuclear and Coulomb interactions. In this approach, the interference is reduced to an additional phase between the Coulomb and nuclear amplitudes. Moreover, several approximations were used in the derivation. First, in order to avoid integrating over off-mass-shell contributions to the nuclear amplitude (essentially unknown), a very slow variation of the nuclear amplitude phase was assumed: $\arg {\cal A}^{\rm N} \approx \hbox{const}$. Then, in order to obtain a closed-form expression, the exponential slope of the nuclear modulus was assumed constant (i.e.~only the $b_1$ parameter is non-zero in the parametrisation Eq.~(\ref{eq:nuc mod})) which is formally incompatible with the existence of the diffractive minimum. The original formula did not contain the electromagnetic form factor ${\cal F}$, which was added later by hand:
\begin{equation}
\label{eq:int swy}
	\begin{aligned}
		{\d\sigma\over \d t}^{\rm C+N} =& {\pi (\hbar c)^2 \over s p^2} \left | {\alpha s\over t} {\cal F}^2 \e^{\I\alpha \Phi(t)} + {\cal A}^{\rm N}
			\right |^2\ ,\cr
		\Phi(t) =& - \left( \log {b_1 |t|\over 2} + \gamma \right)\ ,\cr
	\end{aligned}
\end{equation}
where $\alpha$ is the fine-structure constant and $\gamma \doteq 0.577$ the Euler constant. Despite the many limitations, the formula has been extensively used in past data analyses. For backward-comparison reasons it is also considered in this report.

The approach of {\bf Cahn} \cite{cahn82} uses an impact parameter formalism and is based on the additivity of eikonals. The first part of his derivation does not impose any limit on the nuclear amplitude, leading to the formula (Eq.~(30) in \cite{cahn82}):
\begin{equation}
\label{eq:int cahn}
	\begin{aligned}
		{\d\sigma\over \d t}^{\rm C+N} =& {\pi (\hbar c)^2 \over s p^2} \left | -{\alpha s\over q^2} {\cal F}^2
			+ {\cal A}^{\rm N}\, \Big[1 - \I\alpha G(-q^2)\Big] \right |^2\ ,\cr
		G(-q^2) =& - \int\limits_0^\infty \d q'^2 \log {q'^2\over q^2} {\d\phantom{q'^2}\over \d q'^2} {\cal F}^2(-q^2)\cr
				&+ {1\over\pi} \int\d^2 q' {{\cal F}^2(-q'^2)\over q'^2} \left[ {{\cal A}^{\rm N}\left(-[\vec{q} - \vec{q'}]^2\right) \over
					{\cal A}^{\rm N}(-q^2)} - 1 \right]\ , \cr
	\end{aligned}
\end{equation}
where $t=-q^2$, $\vec{q'}$ is a two-dimensional vector and $q'^2 = |\vec{q'}|^2$. The second part of the article gives simplified formulae for nuclear amplitudes with purely-exponential modulus and constant phase and is, thus, of limited interest for the present analysis.

{\bf Kundr\' at and Lokaj\' i\v cek (KL)} \cite{kl94} transformed the formula of Cahn, Eq.~(\ref{eq:int cahn}), into a form better suited for practical applications and added the kinematic limits on the momentum transfer:\footnote{%
Note that some recent publications by the same authors (e.g.~\cite{kl05,kklp11}) contain a misprint: the wrong sign in front of the second term contributing to $G(t)$.
}
\begin{equation}
\label{eq:int kl}
	\begin{aligned}
		{\d\sigma\over \d t}^{\rm C+N} =& {\pi (\hbar c)^2 \over s p^2} \left | {\alpha s\over t} {\cal F}^2
			+ {\cal A}^{\rm N}\, \Big[1 - \I\alpha G(t)\Big] \right |^2\ ,\cr
		G(t) =& \int\limits_{-4p^2}^0 \d t'\, \log {t'\over t} {\d\phantom{t'}\over \d t'} {\cal F}^2(t')\cr
			  &- \int\limits_{-4p^2}^0 \d t' \left( {{\cal A}^{\rm N}(t') \over {\cal A}^{\rm N}(t)} - 1 \right) { I(t, t')\over 2\pi }\ , \cr
		I(t, t') =& \int_0^{2\pi} \d\phi\ {{\cal F}^2(t'')\over t''}\ ,\cr
		t'' =& t + t' + 2\sqrt{t\, t'} \cos\phi\ .\cr
	\end{aligned}
\end{equation}
A slightly different variant proposed in Eq.~(22) in~\cite{kl05} was considered, too:
\begin{equation}
\label{eq:int kl exp}
	{\d\sigma\over \d t}^{\rm C+N} = {\pi (\hbar c)^2 \over s p^2} \left | {\alpha s\over t} {\cal F}^2
		+ {\cal A}^{\rm N}\, \e^{- \I\alpha G(t)} \right |^2 \ .
\end{equation}

The interference formula by Cahn, Eq.~(\ref{eq:int cahn}), and the KL formula, Eq.~(\ref{eq:int kl}), are very similar by construction and therefore they give practically identical interference effects.

Since the quantities $G$ in Eqs.~(\ref{eq:int cahn}) and (\ref{eq:int kl}) are complex, the interference effects in these treatments are generally more feature-rich than with the SWY formula, Eq.~(\ref{eq:int swy}), where the interference is reduced to a single additional phase $\Phi$.

By analysing Eqs.~(\ref{eq:int swy}), (\ref{eq:int cahn}) and (\ref{eq:int kl}), one can conclude that in the region where the nuclear amplitude dominates ($|t| \gtrsim 0.003\un{GeV^2}$), the effects due to the Coulomb interaction are of the order of $\alpha$ or the ratio $|{\cal A}^{\rm C}| / |{\cal A}^{\rm N}|$. In both cases, the magnitude of the interference effects can be expected at a percent level, as shown in Figure~\ref{fig:cni effect}. The figure also shows that the effects at different $|t|$ probe different parts of the nuclear phase: maximum sensitivity to $\rho$ lies at very low $|t|$ while at higher $|t|$ the effects are sensitive to phase values at slightly higher $|t|$. It can also be observed that for the constant, standard and Bailly phase the effects are very similar and rather mild at higher $|t|$. This can be understood from a very limited variation of the phase at low $|t|$, which is the region contributing most to the integral in Eq.~(\ref{eq:int cahn}) or (\ref{eq:int kl}). On the contrary, the higher $|t|$ response to peripheral phases can have various forms, often similar to the deviation of the reconstructed cross-section from pure-exponential, see the top plots in Figures~\ref{fig:fit exp1} and~\ref{fig:fit exp3}.

\subsection{Analysis Procedure}
\label{sec:cni anal proc}

In addition to using the data from Table~\ref{tab:data}, one might consider including the $\beta^* = 90\un{m}$ data \cite{8tev-90m} which benefit from much smaller uncertainties. However, due to the limited\Break reach, $|t| \gtrsim 0.03\un{GeV^2}$, they have essentially no sensitivity to the $\rho$ parameter, cf.~Fig.~\ref{fig:cni effect}. Furthermore, due to possible systematic tensions between the data sets, the inclusion of the $\beta^* = 90\un{m}$ data may have a deteriorating impact on the $\rho$ determination. Therefore, the value of $\rho$ was determined from the $\beta^* = 1000\un{m}$ data only. For other parameters to which both data sets have non-negligible sensitivity (e.g.~$b_i$ in Eq.~(\ref{eq:nuc mod})), both data sets should give compatible results. This was verified for all the fits that will be presented later on. Since the $\beta^* = 90\un{m}$ data yield much lower uncertainties, both data sets have been used for determining all parameters except $\rho$. In practice, a series of two fits was performed:
\begin{itemize}
\item step 1: fit of $\beta^* = 1000\un{m}$ data with $\rho$ free,
\item step 2: fit of $\beta^* = 1000$ and $90\un{m}$ data with $\rho$ fixed from the preceding step.
\end{itemize}

The standard least-squares method was used for all the fits. In particular, minimising
\begin{equation}
\label{eq:chi sq A}
	\begin{aligned}
		\chi^2 &= \Delta^\T \mat V^{-1} \Delta\ ,\quad
			\Delta_i = \left.{\d\sigma\over \d t}\right|_{{\rm bin}\ i} - {\d\sigma^{\rm C+N}\over\d t}\left(t^{\rm rep}_{{\rm bin}\ i}\right)\ ,\cr
		\mat V &= \mat V_{\rm stat} + \mat V_{\rm syst}\ ,\cr
	\end{aligned}
\end{equation}
where $\Delta$ is a vector of differences between the differential cross-section data and a fit function $\d\sigma^{C+N}/\d t$ evaluated at the representative point $t^{\rm rep}$ of each bin~\cite{lafferty94}. The minimisation is repeated several times, and the representative points are updated between iterations. The CNI effects are calculated using the computer code from~\cite{elegent}. The covariance matrix $\mat V$ has two components. The diagonal of $\mat V_{\rm stat}$ contains the statistical uncertainty squared from Table~\ref{tab:data} and from the Table~3 in \cite{8tev-90m}. $\mat V_{\rm syst}$ includes all systematic uncertainty contributions except the normalisation, see Eq.~(\ref{eq:covar mat}) and Eq.~(14) in \cite{8tev-90m}. For improved fit stability, the normalisation uncertainty is not included in the $\chi^2$ definition. Instead, the uncertainty is propagated for each fit parameter. For this purpose, the fit is repeated with $-1\un{\sigma}$, $0\un{\sigma}$ and $+1\un{\sigma}$ biases independently in: global normalisation ($1\un{\sigma} = 4.2\un{\%}$), $90\un{m}$ data normalisation ($0.08\un{\%}$) and $1000\un{m}$ data normalisation ($0.25\un{\%}$). This gives a sample of 27 fit results, from which one can estimate the propagated normalisation uncertainty of a parameter as $(\hbox{max} - \hbox{min})/2$, where ``max'' (``min'') is the greatest (smallest) value in the sample. This normalisation uncertainty is, at the end, added quadratically to the uncertainty reported by the fit with no bias.

The fits have shown low sensitivity to several of the\Break choices presented above, summarised in the following list. 
\begin{itemize}
\item Choice of the form factor in Eq.~(\ref{eq:coul cs}). The options considered in \cite{elegent} have been tested, none of them giving any significant difference with respect to the default choice \cite{puckett10}.
\item Extension of the modulus of the nuclear amplitude to the unobserved $|t|$ region, see the last paragraph in Section~\ref{sec:cni nuclear modulus}. No effect was observed when the high-$|t|$ part was altered (both shape and normalisation) nor when the size of the transition region was changed.
\item Use of the Cahn or KL formula. Only the latter will be used in what follows to represent both of them.
\item The two variants of the KL formula, Eqs.~(\ref{eq:int kl}) and (\ref{eq:int kl exp}). The latter will be used below.
\item Fits with constant, standard and Bailly phase are practically indistinguishable. This can be expected from\Break Fig.~\ref{fig:cni effect} showing that the corresponding CNI effects are very similar. Therefore, in the remainder of this article, these phases will be treated as a single family represented by the constant phase.
\end{itemize}

One of the goals of this study is to probe the origin of the differential cross-section non-exponentiality reported earlier~\cite{8tev-90m}. Therefore, the following two classes of fits were considered.
\begin{itemize}
\item Section \ref{sec:fit exp1}: fits with purely exponential nuclear modulus, that is $N_b=1$ in Eq.~(\ref{eq:nuc mod}). In this case, the non-exponentiality can come from the CNI effects only.
\item Section \ref{sec:fit exp3}: fits with nuclear modulus flexible enough to describe the non-exponentiality without the CNI effects. Here, the non-exponentiality may be due to the nuclear modulus, CNI effects or both.
\end{itemize}
For each of these nuclear modulus cases, the following two phase parametrisations were considered:
\begin{itemize}\setlength\itemsep{0pt}
\item constant phase, Eq.~(\ref{eq:nuc phase con}), as a representative of the\Break central-phases family,
\item peripheral phase, Eq.~(\ref{eq:nuc phase per}) with parameters fixed to the values in Eq.~(\ref{eq:nuc phase per val}) to represent peripheral behaviour in the impact parameter space.
\end{itemize}

In each case, the fit results are used to calculate the total cross-section via the optical theorem:
\begin{equation}
\label{eq:si tot}
\sigma_{\rm tot}^2 = {16\pi\, (\hbar c)^2\over 1 + \rho^2}\, a\ .
\end{equation}
Note that unlike all previous total cross-section determinations at LHC, in this article all the ingredients come consistently from a single analysis.

\subsection{Fits with Purely Exponential Nuclear Modulus}
\label{sec:fit exp1}

\begin{figure*}
\begin{center}
\includegraphics{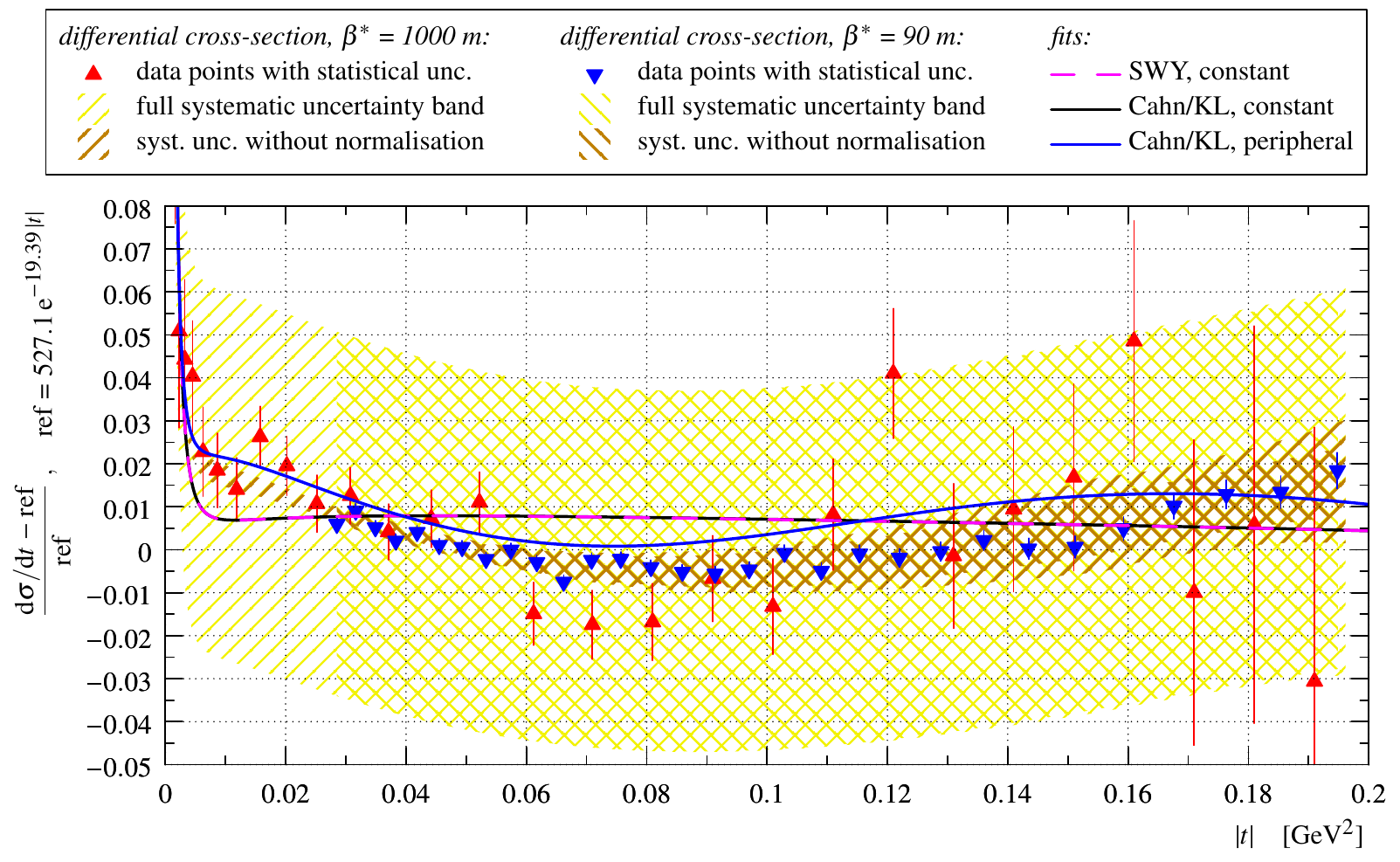}
\includegraphics{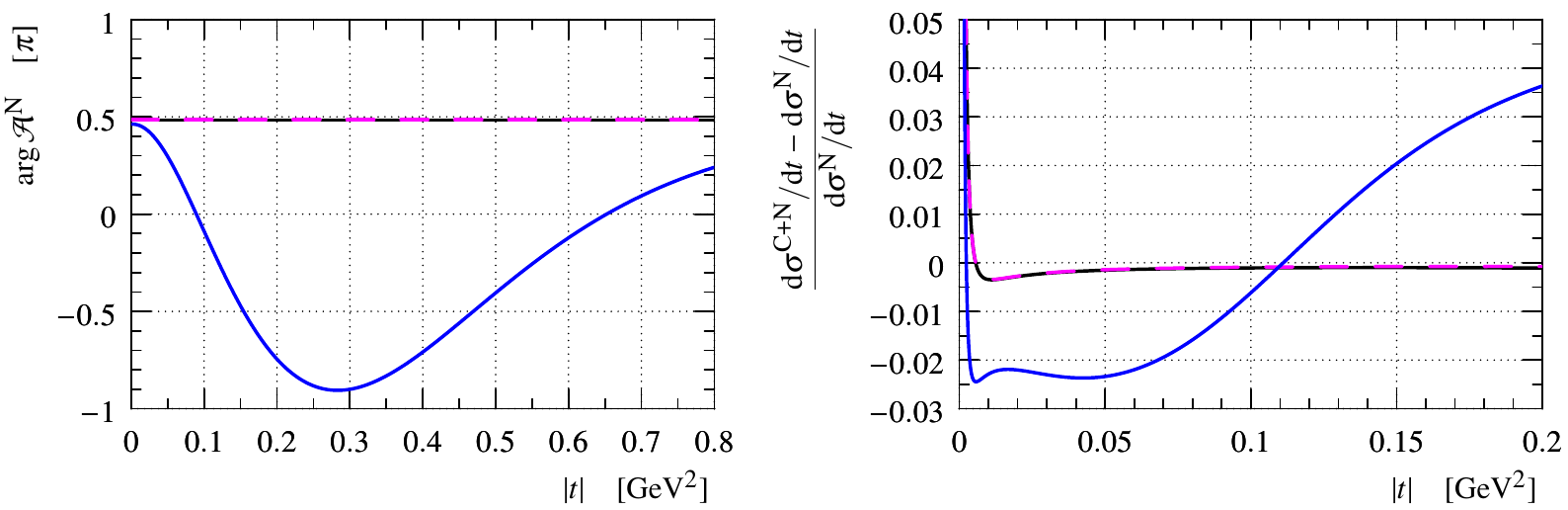}
\caption{Visualisation of the fit results from Table~\ref{tab:fit exp1} obtained with $N_b=1$. The continuous (dashed) lines correspond to fits with Cahn or KL (SWY) formula. Note that the fits with constant nuclear phase largely overlap.
TOP: fits compared to differential cross-section data in a relative reference frame, see the vertical axis label. The reference is identical to the one in \cite{8tev-90m}. 
BOTTOM LEFT: $t$-dependence of the nuclear phase as extracted from the fits.
BOTTOM RIGHT: the effects induced by the Coulomb interaction for each of the fits.
}%
\label{fig:fit exp1}
\end{center}
\end{figure*}

The goal of this section is to test whether the data are compatible with a purely exponential nuclear modulus, i.e.~$N_b=1$ in Eq.~(\ref{eq:nuc mod}). In other words, the non-exponentiality is\Break forced to originate from the Coulomb-induced effects. The fit results obtained with the KL and (where applicable) SWY formulae are summarised in Table~\ref{tab:fit exp1} and graphically shown in Fig.~\ref{fig:fit exp1}.

\begin{table*}
\caption{Fit results with $N_b=1$. Each column corresponds to a fit with different interference formula and/or nuclear phase.}
\vskip-2mm
\label{tab:fit exp1}
\begin{center}
\setlength\tabcolsep{2.5mm}
\input fig/fit_exp1/table_data.tex
\end{center}
\end{table*}

Table~\ref{tab:fit exp1} shows that both fits with constant phase are essentially identical and have bad quality. The step-2 fit using both $\beta^*=1000$\,m and $90\un{m}$ data can be excluded with $7.6\un{\sigma}$ significance. Consequently, since the combination of $N_b=1$ and constant phase is the only one compatible with the SWY approach, that formula is experimentally excluded even on the basis of only the low-$|t|$ data set discussed here. This result is complementary to the observation of a diffractive minimum at $\sqrt{s} = 8\un{TeV}$ (to be published in a forthcoming article) which also contradicts the assumptions of the SWY formula.

Although the quality of the fit with the peripheral phase is good, this option seems disfavoured from different perspectives.
\begin{itemize}
\item There are several theoretical reasons for the nuclear component not to be purely exponential, e.g.~\cite{kmr00-multiPom,kmr12-opacity,kmr15-slope,fagundes15}. Indeed, most elastic scattering models predict\Break a non-exponential nuclear modulus, see e.g.~\cite{elegent} and references therein.
\item The value of $\rho$ obtained in this fit may be regarded as an outlier with respect to a consistent pattern of other fits from this article and extrapolations from lower energies: e.g.~\cite{fagundes12,block12,compete} and most models in \cite{elegent}.
\end{itemize}
Let us also recall that the good quality of this fit is possible due to the more complex KL formula where the CNI effects go beyond a simple additional phase in the traditional SWY concept.

\subsection{Fits with Non-Exponential Nuclear Modulus}
\label{sec:fit exp3}

The aim of this section is to discuss fits with enough flexibility in the nuclear modulus to describe the non-exponentiality in the data. Since a non-exponential hadronic modulus is used, the only applicable interference formula is KL. $N_b=2$ to $5$ were considered. The optimal degree was chosen according to two criteria: reasonable $\chi^2/\hbox{ndf}$ and stability of fit parameters (among which $\rho$ is one of the most sensitive). For instance, with constant phase the fit (step 1) with $N_b=2$ yields $\chi^2/\hbox{ndf} = 1.07$ and $\rho = 0.10$ while the one with $N_b=3$ gives $\chi^2/\hbox{ndf} = 1.03$ and $\rho = 0.12$. Both fits have the normalised $\chi^2$ reasonably close to $1$, but the value of $\rho$ changes significantly between $N_b=2$ and $3$ which is unexpected should $N_b=2$ be sufficient. On the other hand $N_b=4$ gives $\chi^2/\hbox{ndf} = 0.861$ which is unreasonably low. Therefore $N_b=3$ was chosen.

\begin{figure}
\begin{center}
\includegraphics{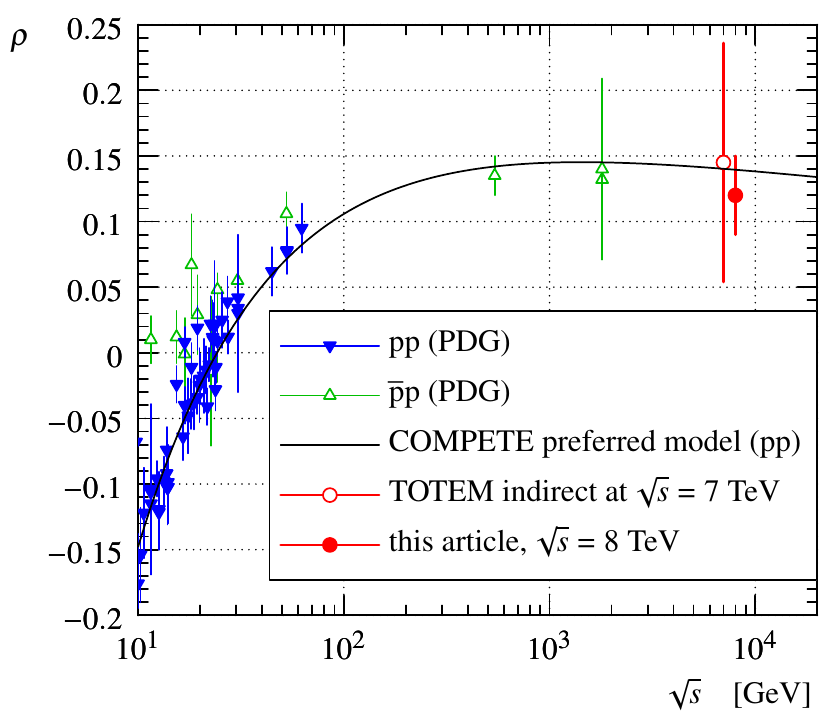}
\caption{Energy dependence of the $\rho$ parameter. The blue (green) triangles correspond to $\rm pp$ ($\rm \bar pp$) data from PDG \cite{pdg} -- note that most of these points were determined with the help of the SWY formula, shown to be inconsistent with the present data. The hollow red circle stands for the earlier indirect determination by TOTEM~\cite{epl101-tot}. The filled red circle represents the two results from Table~\ref{tab:fit exp3} which are numerically identical within the resolution. The black curve gives the preferred $\rm pp$ model by COMPETE~\cite{compete}, obtained without using LHC data.
}%
\label{fig:rho cmp exp3}
\end{center}
\end{figure}

As shown in Table~\ref{tab:fit exp3}, both fits have reasonable fit quality and remarkably consistent values of $\rho$ (identical within the resolution) which are compared to previous determinations at lower energies in Fig.~\ref{fig:rho cmp exp3}. Take note that the obtained parameters for the nuclear amplitude ($a$ and $b_i$) are consistent between step 1 ($\beta^* = 1000\un{m}$ data only) and step 2 (both $\beta^* = 1000$ and $90\un{m}$ data) of the fitting procedure as already mentioned in Section~\ref{sec:cni anal proc}.

\begin{figure*}
\begin{center}
\includegraphics{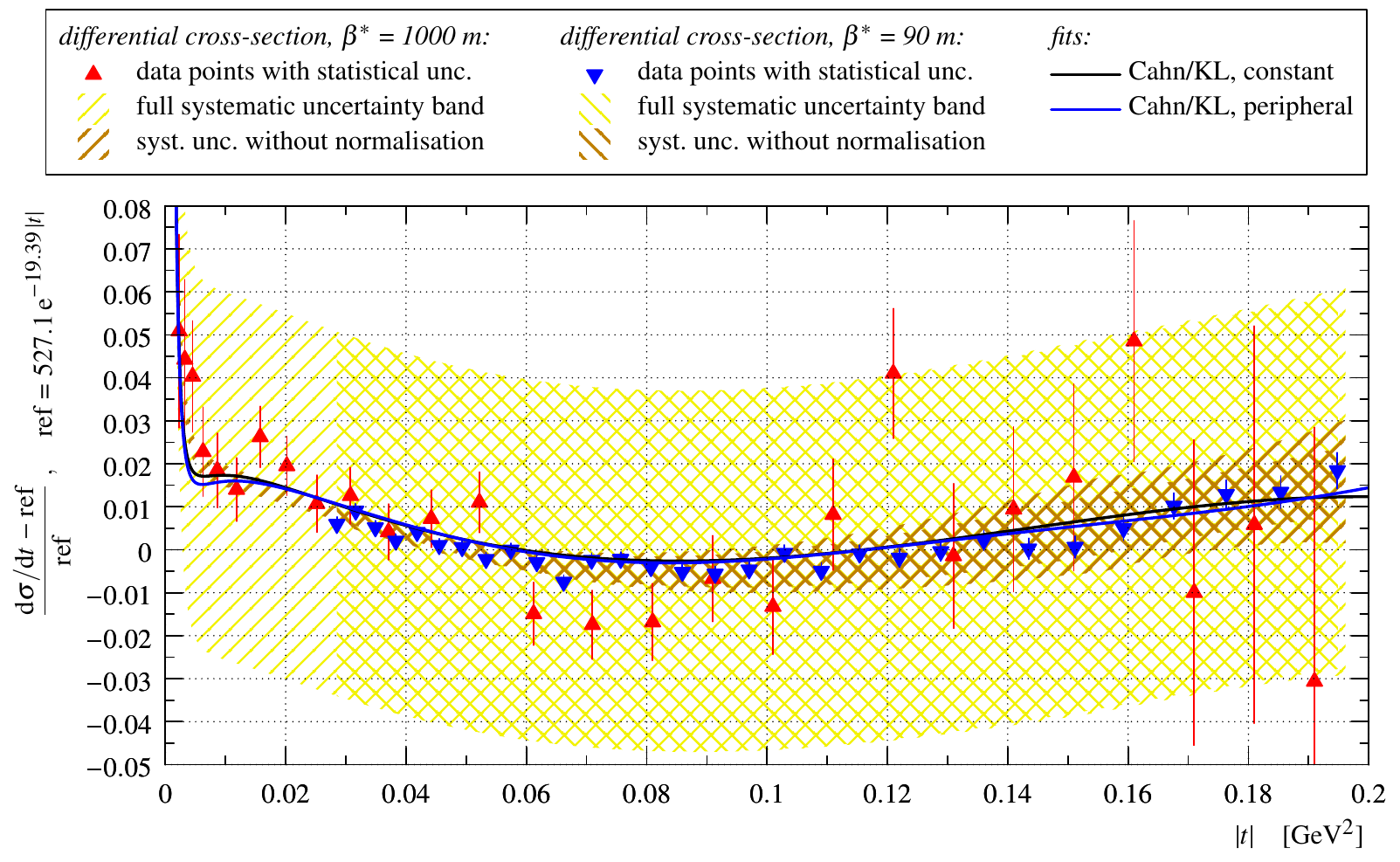}
\includegraphics{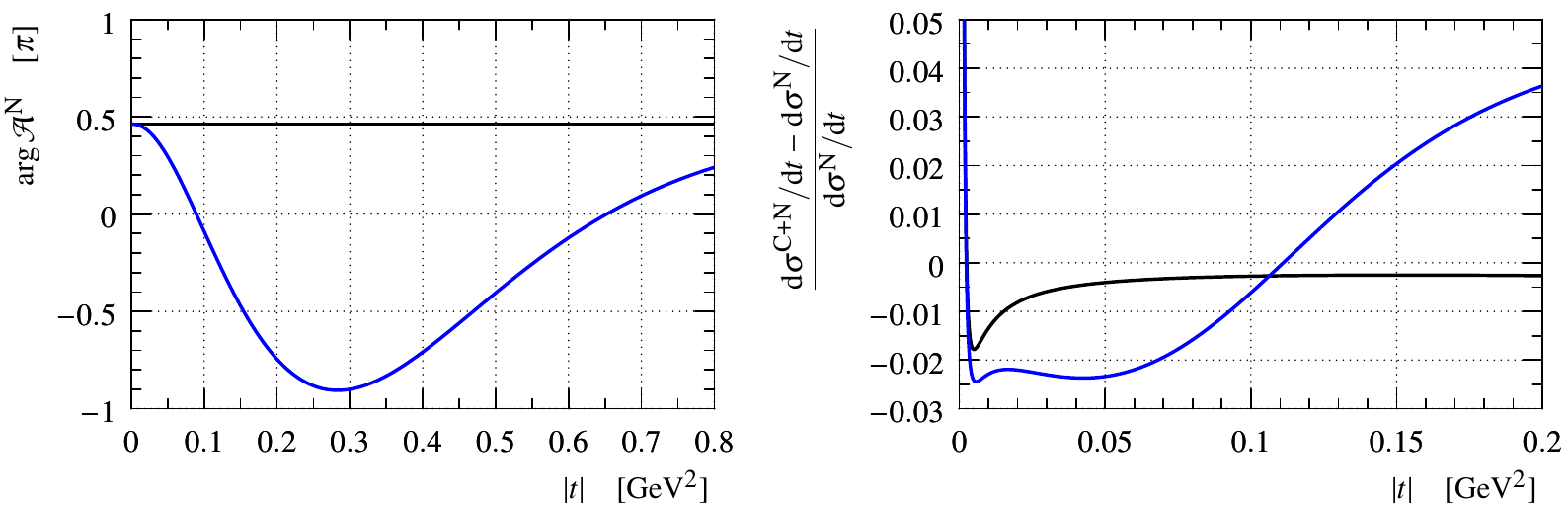}
\caption{Visualisation of the fit results from Table~\ref{tab:fit exp3} obtained with Cahn or KL formula and $N_b=3$. The solid lines correspond to fits with different nuclear phases.
TOP: fits compared to differential cross-section data in a relative reference frame, see the vertical axis label. The reference is identical to the one in \cite{8tev-90m}. 
BOTTOM LEFT: $t$-dependence of the nuclear phase as extracted from the fits.
BOTTOM RIGHT: the effects induced by the Coulomb interaction for each of the fits.
}%
\label{fig:fit exp3}
\end{center}
\end{figure*}

Fig.~\ref{fig:fit exp3} shows that the level of Coulomb-induced effects is very different in the fits. It is much stronger in the case of the peripheral-phase, which can be expected as this phase features a faster variation in the low-$|t|$ region.

\begin{table*}
\caption{Fit results with Cahn or KL formula and $N_b=3$.}
\vskip-2mm
\label{tab:fit exp3}
\begin{center}
\setlength\tabcolsep{5mm}
\input fig/fit_exp3/table_data.tex
\end{center}
\end{table*}

The total cross-section results from the two fits in Table~\ref{tab:fit exp3} are well consistent with each other and also with previous measurements \cite{8tev-90m,prl111}. The slightly higher values with respect to previous analyses neglecting the Coulomb interaction are expected as long as $\rho > 0$. This gives negative interference at low $|t|$ and when separated leads to an increase of nuclear cross-section intercept $a$ and thus also total cross-section via Eq.~(\ref{eq:si tot}).

\begin{figure}
\begin{center}
\includegraphics{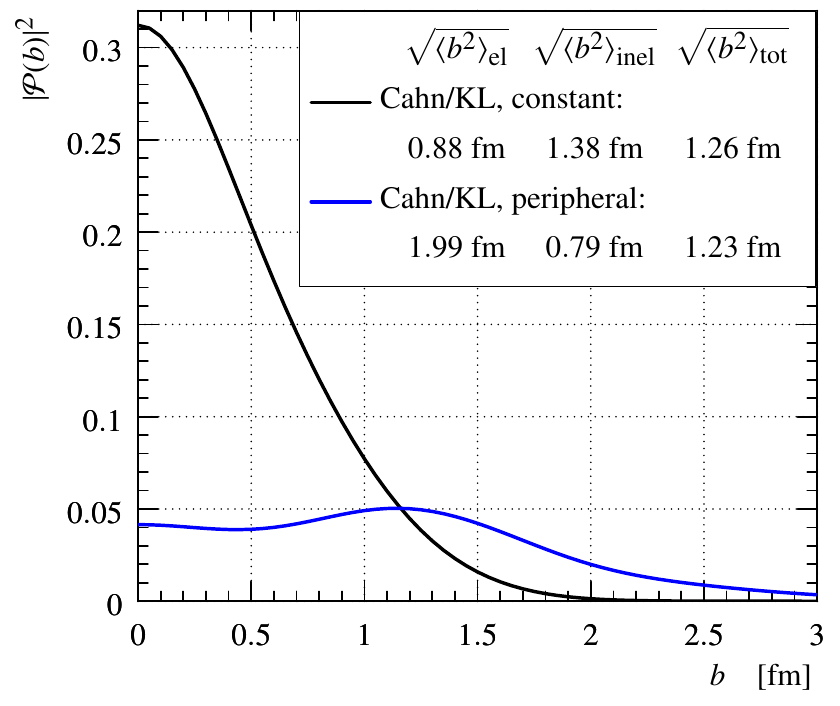}
\caption{%
Square of the impact-parameter amplitude, ${\cal P}$, as a function of impact parameter, $b$. The two lines correspond to the fits in Table~\ref{tab:fit exp3}, using the same colour code as in Figure~\ref{fig:fit exp3}. The root-mean-squares of $b$ in the legend are calculated from Eq.~(\ref{eq:ms b}).
}%
\label{fig:bdist exp3}
\end{center}
\end{figure}

It is interesting to study the fit behaviour in the impact-parameter space. The scattering amplitude in this representation (sometimes called profile function), ${\cal P}(b)$, can be obtained from the nuclear amplitude by means of\Break Fourier-Bessel transformation (see e.g.~\cite{klk02}):
\begin{equation}
\label{eq:prof fun}
	\begin{aligned}
		&{\cal P}(b) = {1\over 4 p \sqrt s} \int\limits_{-\infty}^0 \d t\,J_0\left(b\sqrt{-t}\over \hbar c\right)\,{\cal A}^{\rm N}(t)\ ,\cr
		&\hbox{normalised that } {\sigma_{\rm el} = 8\pi \int\limits_0^{+\infty} b\, \d b\, |{\cal P}(b)|^2}\ ,\cr
	\end{aligned}
\end{equation}
where $\sigma_{\rm el}$ is the integrated elastic cross-section. The profile functions for the two fits from Table~\ref{tab:fit exp3} are shown in Figure~\ref{fig:bdist exp3}. The fit with constant nuclear phase gives a distribution peaked at $b=0$. It corresponds to a behaviour that is more central than for the fit with peripheral phase, where the amplitude modulus reaches maximum at $b\approx 1.2\un{fm}$. These considerations can be extended to inelastic channels. Following Section~3 in~\cite{klk02}, one can calculate the mean values of $b^2$ for elastic ($\langle b^2\rangle_{\rm el}$), inelastic ($\langle b^2\rangle_{\rm inel}$) or all ($\langle b^2\rangle_{\rm tot}$) collisions:
\begin{equation}
\label{eq:ms b}
	\begin{aligned}
		\langle b^2\rangle_j &= {
			\int b\,\d b\,b^2\, h_j(b)
			\over
			\int b\,\d b\, h_j(b)
		}\ ,\qquad h_{\rm el}(b) = |{\cal P}(b)|^2\cr
		h_{\rm tot}(b) &= \Im {\cal P}(b)\ ,\qquad
		h_{\rm inel}(b) = h_{\rm tot}(b) - h_{\rm el}(b)\ .\cr
	\end{aligned}
\end{equation}
Their values reproduced in Figure~\ref{fig:bdist exp3} indicate that the fit with constant nuclear phase leads to a picture with elastic collisions more central than the inelastic ones. The hierarchy is inverted for the fit with peripheral phase.

%% file: fig/fit_exp1/table_data.tex
\begin{tabular}{lcccc}\hline
										& SWY, constant 		& Cahn/KL, constant		& Cahn/KL, peripheral	\cr\hline
$\hbox{step 1:}\, \chi^2/\hbox{ndf}$	& $ 48.0 / 27 = 1.78$	& $ 48.1 / 27 = 1.78$	& $ 27.7 / 27 = 1.03$	\cr
$\hbox{step 2:}\, \chi^2/\hbox{ndf}$	& $ 180.8 / 58 = 3.12$	& $ 181.2 / 58 = 3.12$	& $ 64.3 / 58 = 1.11$	\cr
\hline
$a\ung{mb/GeV^2}$				 		& $533 \pm 23$			& $533 \pm 23$			& $551 \pm 23$			\cr
$b_1\ung{GeV^{-2}}$				 		& $19.42 \pm  0.05$ 	& $19.42 \pm  0.05$		& $19.74 \pm  0.05$		\cr
\hline
$\rho$							 		& $0.05 \pm  0.02$  	& $0.05 \pm  0.02$		& $0.10 \pm  0.02$		\cr
$\zeta_1$						 		&					 	&					 	& $800$					\cr
$\kappa$						 		&					 	&					 	& $2.311$				\cr
$\nu\ung{GeV^{-2}}$				 		&					 	&					 	& $8.161$				\cr
\hline
$\sigma_{\rm tot}\ung{mb}$		 		& $102.0 \pm  2.2$		& $102.0 \pm  2.2$		& $103.4 \pm  2.3$		\cr
\hline
\end{tabular}

%% file: fig/fit_exp3/table_data.tex
\begin{tabular}{lccc}\hline
										& Cahn/KL, constant		& Cahn/KL, peripheral	\cr\hline
$\hbox{step 1:}\, \chi^2/\hbox{ndf}$	& $ 25.7/ 25 = 1.03$	& $ 25.0/ 25 = 1.00$	\cr
$\hbox{step 2:}\, \chi^2/\hbox{ndf}$	& $ 57.5/ 56 = 1.03$	& $ 57.6/ 56 = 1.03$	\cr
\hline
$a\ung{mb/GeV^2}$						& $549 \pm 24$			& $549 \pm 24$			\cr
$b_1\ung{GeV^{-2}}$						& $20.47 \pm  0.14$		& $19.56 \pm  0.13$		\cr
$b_2\ung{GeV^{-4}}$						& $8.8 \pm  1.6$		& $-3.3 \pm  1.5$		\cr
$b_3\ung{GeV^{-6}}$						& $20 \pm  6$			& $-13 \pm  5$			\cr
\hline
$\rho$									& $0.12 \pm  0.03$		& $0.12 \pm 0.03$		\cr
$\zeta_1$								&					 	& $800$			   		\cr
$\kappa$								&					 	& $2.311$		   		\cr
$\nu\ung{GeV^{-2}}$						&					 	& $8.161$		   		\cr
\hline
$\sigma_{\rm tot}\ung{mb}$				& $102.9 \pm  2.3$		& $103.0 \pm  2.3$		\cr
\hline
\end{tabular}

%% file: conclusions.tex
\section{Summary and Outlook}
\label{sec:summary}
For the first time at LHC the differential cross-section of elastic proton-proton scattering has been measured at $|t|$-values down to the Coulomb-nuclear interference (CNI) region. This was made possible by a special beam optics, a novel collimation procedure and by moving the RPs to an unprecedented distance of only $3\,\sigma$ from the centre of the circulating beam.

To fit $\d\sigma/\d t$ in the CNI region, several interference formulae -- Simplified West and Yennie (SWY), Cahn and\Break Kundr\' at-Lokaj\' i\v cek (KL) -- were explored in conjunction\Break with different mathematical descriptions of the modulus and phase of the nuclear amplitude as a function of $t$. The nuclear modulus was parametrised as an exponential function with a polynomial of degree $N_b=1$ or $3$ in the exponent. These two alternatives allowed to test whether the nuclear modulus can be purely exponential or more flexibility is required. For the phase two options were considered, leading to different impact-parameter distributions of elastic scattering events: a constant phase implying a central behaviour, and another description favouring peripheral collisions. The following conclusions can be drawn.
\begin{itemize}
\item Purely exponential nuclear modulus ($N_b=1$), constant phase: excluded with more than $7\un{\sigma}$ confidence. Since this is the only combination compatible with the SWY formula, the data exclude the usage of the formula.
\item Purely exponential nuclear modulus ($N_b=1$), peripheral phase: the data do not exclude this option which, however, is disfavoured from other perspectives.
\item Non-exponential nuclear modulus ($N_b=3$): both constant and peripheral phases are well compatible with the data, therefore the central impact-parameter picture\Break prevalent in phenomenological descriptions is not a necessity.
\end{itemize}

The $\rho$ parameter was for the first time at LHC extracted via the Coulomb-nuclear interference. In the preferred fits ($N_b=3$):
\begin{equation}
\label{eq:rho final}
\rho = 0.12 \pm 0.03\ .
\end{equation}

The new total cross-section determination is conceptually more accurate than in all previous LHC publications since the CNI effects are explicitly treated. Moreover, the value of $\rho$ comes from the same analysis, not from an external source, which underlines consistency. The $\sigma_{\rm tot}$ values are very well consistent among all non-excluded fits and compatible with the previous measurements. As expected, the new determination yields slightly greater values relative to previous results where the negative CNI was not taken into account. Also note that if the SWY formula with purely exponential hadronic modulus is used, the total cross-section is underestimated by about $1\un{mb}$. A similar underestimation may occur if the non-exponentiality is not taken into account~\cite{block16}.

For even stronger results in the future the key point is a better distinction between the nuclear and CNI cross-section components, which can be achieved from both theoretical and experimental sides. New theory developments may narrow down the range of allowed parametrisations of the nuclear modulus and phase or better constrain the induced CNI effects. The experimental improvements include increasing statistics and reducing the lower $|t|$ threshold. For the former, TOTEM has already upgraded the RP mechanics such that both vertical pots can be simultaneously placed very close to the beam. For the latter, TOTEM foresees an optics with extremely high $\beta^* \approx 2500\un{m}$ which would allow to reach the CNI region even at Run II energies. Moreover, recent experience with the $\beta^* = 90\un{m}$ optics at $\sqrt s = 13\un{TeV}$ shows that very low beam emittances can be achieved, thus possibly further reducing the RP distance from the beam.

%% file: acknowledgements.tex
This work was supported by the institutions\Break listed on the front page and also by the 
Magnus Ehrnrooth foundation (Finland), the Waldemar von Frenckell foundation (Finland), 
the Academy of Finland, the Finnish Academy of Science and Letters (The Vilho, Yrj\"o and Kalle V\"ais\"al\"a Fund), 
the OTKA grant NK 101438 (Hungary). Individuals have received support from Nylands nation vid Helsingfors universitet (Finland) 
and from the M\v SMT \v CR (Czech Republic).

%% file: elastic_1km_epjc.bbl
\begin{thebibliography}{99}
\bibitem{plb43}
	\Name{U.~Amaldi et al.}
	\Review{Phys.~Lett.~B}{43}{1973}{231 -- 236}.

\bibitem{plb66}
	\Name{U.~Amaldi et al.}
	\Review{Phys.~Lett.~B}{66}{1977}{390 -- 394}.

\bibitem{npb141}
	\Name{L.~Baksay et al.}
	\Review{Nucl.~Phys.~B}{141}{1978}{1 -- 28}.

\bibitem{prl47}
	\Name{D.~Favart et al.}
	\Review{Phys.~Rev.~Lett.}{47}{1981}{1191 -- 1194}.

\bibitem{plb115}
	\Name{M.~Ambrosio et al.}
	\Review{Phys.~Lett.~B}{115}{1982}{495 -- 502}.

\bibitem{plb120}
	\Name{N.~Amos et al.}
	\Review{Phys.~Lett.~B}{120}{1983}{460 -- 464}.

\bibitem{plb128}
	\Name{N.~Amos et al.}
	\Review{Phys.~Lett.~B}{128}{1983}{343 -- 348}.

\bibitem{npb262}
	\Name{N.~Amos et al.}
	\Review{Nucl.~Phys.~B}{262}{1985}{689 -- 714}.

\bibitem{plb198}
	\Name{D.~Bernard et al.}
	\Review{Phys.~Lett.~B}{198}{1987}{583 -- 589}.

\bibitem{plb316}
	\Name{C.~Augier et al.}
	\Review{Phys.~Lett.~B}{316}{1993}{448 -- 454}.

\bibitem{prl68}
	\Name{N.~A.~Amos et al.}
	\Review{Phys.~Rev.~Lett.}{68}{1992}{2433 -- 2436}.

\bibitem{wy68}
	\Name{G.~B.~West and D.~R.~Yennie}
	\Review{Phys. Rev.}{172}{1968}{1413 -- 1422}.

\bibitem{dremin-dispersion}
	\Name{I.~M.~Dremin}
	\Review{JETP Lett.}{97}{2013}{571 -- 573},
	arXiv:1304.5345; and references therein.

\bibitem{8tev-90m}
	\Name{G.~Antchev~\etal{}~(TOTEM Collaboration)}
	\Review{Nucl.~Phys.~B}{899}{2015}{527 -- 546}.

\bibitem{cahn82}
	\Name{R.~Cahn}
	\Review{Z.~Phys.~C}{15}{1982}{253 -- 260}.

\bibitem{kl94}
	\Name{V.~Kundr\' at and M.~Lokaj\' i\v cek}
	\Review{Z.~Phys.}{C63}{1994}{619--630}.

\bibitem{totem-jinst}
	\Name{G.~Anelli \etal{}~(TOTEM Collaboration)}
	\Review{JINST}{3}{2008}{S08007}.

\bibitem{totem-ijmp}
	\Name{G.~Antchev \etal{}~(TOTEM Collaboration)}
	\Review{Int.~J.~Mod.~Phys.~A}{28}{2013}{1330046}.

\bibitem{epl96}
	\Name{G.~Antchev \etal{}~(TOTEM Collaboration)}
	\Review{Europhys.~Lett.}{96}{2011}{21002}.

\bibitem{epl101-el}
	\Name{G.~Antchev \etal{}~(TOTEM Collaboration)}
	\Review{Europhys.~Lett.}{101}{2013}{21002}.

\bibitem{epl101-tot}
	\Name{G.~Antchev \etal{}~(TOTEM Collaboration)}
	\Review{Europhys.~Lett.}{101}{2013}{21004}.

\bibitem{prl111}
	\Name{G.~Antchev \etal{}~(TOTEM Collaboration)}
	\Review{Phys.~Rev.~Lett.}{111}{2013}{012001}.

\bibitem{totem-optics}
	\Name{G.~Antchev \etal{}~(TOTEM Collaboration)}
	\Review{New J.~Phys.}{16}{2014}{103041}.

\bibitem{op-elog} CERN LHC OP eLogbook:\\
	\url{http://elogbook.cern.ch/eLogbook/eLogbook.jsp?shiftId=1049006},\\
	\url{http://elogbook.cern.ch/eLogbook/eLogbook.jsp?shiftId=1049013},\\ 
	\url{http://elogbook.cern.ch/eLogbook/eLogbook.jsp?shiftId=1049020}.

\bibitem{hubert-thesis}
	\Name{H.~Niewiadomski}
    PhD thesis, CERN-THESIS-2008-080,
	\url{http://cds.cern.ch/record/1131825}.

\bibitem{todesco-lpc}
	\Name{E.~Todesco}
	presentation in the LPC Meeting on 27 June 2016,
	\url{https://indico.cern.ch/event/541101/contributions/2197629/attachments/1298782/1937817/et\_1606\_energy.pdf}

\bibitem{lafferty94}
	\Name{G.~D.~Lafferty and T.~R.~Wyatt}
	\Review{Nucl.~Instrum.~Meth.}{A 355}{1995}{541}.

\bibitem{block96}
	\Name{M.~M.~Block}
	\Review{Phys.~Rev.~D}{54}{1996}{4337 -- 4343}.
	
\bibitem{jan_thesis}
	\Name{J.~Ka\v spar}
	PhD Thesis, CERN-THESIS-2011-214,
	\url{http://cdsweb.cern.ch/record/1441140}.

\bibitem{kklp11}
	\Name{J.~Ka\v spar, V.~Kundr\' at, M.~V.~Lokaj\' i\v cek and J.~Proch\' azka}
	\Review{Nucl.~Phys.~B}{843}{2011}{84 -- 106}.

\bibitem{elegent}
	\Name{J.~Ka\v spar}
	\Review{Comp.~Phys.~Comm.}{185}{2014}{1081 -- 1084}
	and \url{http://elegent.hepforge.org/}.

\bibitem{epl95}
	\Name{G.~Antchev \etal{}~(TOTEM Collaboration)}
	\Review{Europhys.~Lett.}{95}{2011}{41001}.


\bibitem{bailly87}
	\Name{J.~L.~Bailly \etal{}~(EHS-RCBC Collaboration)}
	\Review{Z.~Phys.~C}{37}{1987}{7 -- 16}.

\bibitem{pk16}
	\Name{J.~Proch\'azka and V.~Kundr\'at}
	Eikonal model analysis of elastic hadron collisions at high energies,
	arXiv:1606.09479 (2016).

\bibitem{klk02}
	\Name{V.~Kundr\' at, M.~Lokaj\' i\v cek and D.~Krupa}
	\Review{Phys.~Lett.~B}{544}{2002}{132 -- 138}.

\bibitem{kl96}
	\Name{V.~Kundr\' at and M.~Lokaj\' i\v cek}
	\Review{Mod.~Phys.~Lett.~A}{11}{1996}{2241 -- 2250}.

\bibitem{kl05}
	\Name{V.~Kundr\' at and M. Lokaj\' i\v cek}
	\Review{Phys.~Lett.}{B611}{2005}{102 -- 110}.

\bibitem{puckett10}
	\Name{A.~J.~R.~Puckett \etal{}~(GEp-III Collaboration)}
	Final Results of the GEp-III Experiment and the Status of the Proton Form Factors,
	arXiv:nucl-ex/1008.0855 (2010).

\bibitem{kmr00-multiPom}
	\Name{V.~A.~Khoze, A.~D.~Martin and M.~G.~Ryskin}
	\Review{Eur.~Phys.~J.~C}{18}{2000}{167 -- 179}.

\bibitem{kmr12-opacity}
	\Name{M.~G.~Ryskin, A.~D.~Martin and V.~A.~Khoze}
	\Review{Eur.~Phys.~J.~C}{72}{2012}{1937}.

\bibitem{kmr15-slope}
	\Name{A.~D.~Martin, V.~A.~Khoze and M.~G.~Ryskin}
	\Review{J.~Phys.}{G42}{2015}{2, 025003}.

\bibitem{fagundes15}
	\Name{D.~A.~Fagundes \etal{}}
	Fine structure of the diffraction cone: from ISR to the LHC,
	arXiv:hep-ph/1509.02197 (2015).

\bibitem{fagundes12}
	\Name{D.~A.~Fagundes, E.~G.~S.~Luna, M.~J.~Menon and A.~A.~Natale}
	\Review{Nucl.~Phys.~A}{886}{2012}{40 -- 70}

\bibitem{block12}
	\Name{M.~M.~Block and F.~Halzen}
	\Review{Phys.~Rev.~D}{86}{2012}{014006}.

\bibitem{compete} 
	\Name{J.~R.~Cudell \etal{}~(COMPETE Collaboration)}
	\Review{Phys.\ Rev.\ Lett.}{89}{2002}{201801}.

\bibitem{pdg} 
	\Name{K.~Nakamura \etal{}~(Particle Data Group)}
	\Review{J.~Phys.}{G37}{2010}{075021}.

\bibitem{block16}
	\Name{M.~M.~Block, L.~Durand, P.~Ha and F~.Halzen}
	\Review{Phys.~Rev.~D}{93}{2016}{114009}.

\end{thebibliography}
